\documentclass[aps, prr, twocolumn, amssymb, amsmath, amsfonts, showpacs, superscriptaddress, preprintnumbers, a4paper]{revtex4-2} 
\usepackage{amssymb, amsmath, amsthm}
\usepackage{xcolor}
\usepackage{graphicx} % Required for inserting images
\usepackage{hyperref}
\usepackage{comment}
\usepackage{physics}
\usepackage{footnote}

\begin{document}

\title{Finite-temperature effects on the single-particle Green's function of a mobile impurity in a Luttinger liquid}

\author{Lorenzo Gotta}
\email{lorenzo.gotta@unige.ch}
\affiliation{Department of Quantum Matter Physics, University of Geneva, 24 Quai Ernest-Ansermet, 1211 Geneva, Switzerland}

\author{Thierry Giamarchi}
\affiliation{Department of Quantum Matter Physics, University of Geneva, 24 Quai Ernest-Ansermet, 1211 Geneva, Switzerland}

\date{\today}

\begin{abstract} 
We consider the problem of a quadratically-dispersing quantum impurity interacting through a contact density-density term with a one-dimensional (1D) superfluid described by Luttinger liquid theory. By means of a field-theoretical approach based on the linked-cluster expansion, we quantitatively address the effects of a finite temperature of the 1D bath on the momentum-space single-particle Green's function. Through a careful identification of the resonant phonon-impurity scattering processes, we obtain asymptotic large-time estimates for the thermally-induced exponential damping of the zero-temperature signal both in the \textit{quasiparticle} regime, dominated by the presence of a finite lifetime for the host impurity, and in the \textit{infrared-dominated} regime, where the impurity causes a Anderson's orthogonality catastrophe of the $1D$ bath. In both cases, we identify a temperature-induced coherence time that sets the characteristic timescale of the exponentially-decaying finite-temperature correction to the amplitude of the zero-temperature single-particle Green's function. Moreover, we draw connections between the aforesaid thermal timescale and the microscopic parameters of the Lieb-Liniger model, which admits an effective Luttinger-liquid description, highlighting the dependence of the inverse coherence time on the intra-bath interactions across the different dynamical regimes of the impurity.
\end{abstract}

\maketitle

\section{Introduction}

The problem of a mobile quantum impurity interacting with a one-dimensional many-body environment has attracted sustained interest across condensed matter and cold-atom physics\cite{Zvonarev_2007_Spin,Kantian_2014_regimes,Kamar_2025_ladder,Gotta_2024_Dirac,Lamacraft_2009_spectral,Schecter_2012_velocity,Burovski_2014_relaxation,Shen_2007_transport,Enss_2005_correlation,Sacha_2006_bec,Lesage_1997_correlations,Bonart_2012_brownian}. In one dimension, the host medium is generically described by a Luttinger liquid\cite{Haldane_1981_Luttinger,Giamarchi_2003_1D}, where low-energy excitations are collective and the conventional quasiparticle picture breaks down\cite{giamarchi_book_1d,Bouchoule_2025_Platforms,Schonhammer_2004_Luttinger,Fisher_1997_transport,Pham_2000_fractional}. As a consequence, even a single impurity exhibits highly nontrivial dynamics, shaped by its coupling to gapless bosonic modes.

A central object to characterize these effects is the impurity single-particle Green’s function, which encodes both dynamical and spectral properties. At zero temperature, its behavior is known to display nontrivial power-law decay in time and singular features in frequency within the so called infrared (IR) regime, reflecting the critical nature of the underlying liquid and the absence of well-defined quasiparticles\cite{Neto_1996_heavy,Cheianov_2004_Nonunitary,Cheianov_2004_Zero,Zvonarev_2007_Spin,Massel_2013_dynamics,Kantian_2014_regimes,Kamar_2019_Dynamics,Kamar_2025_ladder,Gotta_2024_Dirac}. These features are closely related to the physics captured by Luttinger liquid theory and to impurity-induced orthogonality effects\cite{Anderson_1967_aoc,Qin_1996_catastrophe,Furusaki_1997_orthogonality,Caracanhas_2015_Vortex}.

Despite this well-established zero-temperature picture, much less is understood about the role of finite temperature, which is unavoidable in experimental realizations\cite{Fukuhara_2013_quantum,Catani_2012_bose,Cavazos_2023_thermal}. Thermal fluctuations introduce an additional energy scale that competes with the intrinsic low-energy dynamics of the system, potentially modifying both the temporal decay and the spectral structure of the impurity Green’s function\cite{Kane_1992_transport,Mattsson_1997_boundaries,Juttner_1997_exact,Juttner_1998_hubbard}. In particular, finite temperature is expected to act as an infrared cutoff, smoothing zero-temperature singularities and inducing finite coherence times for the impurity\cite{Le_Hur_2006_lifetime}.

In this work, we investigate finite-temperature corrections to the single-particle Green’s function of a mobile impurity in a Luttinger liquid. Focusing on the momentum-resolved and time-dependent behavior, we analyze how thermal effects modify the algebraic decay characteristic of the zero-temperature limit, giving rise to a crossover toward thermally dominated dynamics. We show that the interplay between impurity motion and collective excitations leads to a momentum-dependent decoherence and a nontrivial redistribution of spectral weight. Our results provide a unified framework to understand impurity dynamics beyond the zero-temperature regime and offer direct implications for ongoing experiments in one-dimensional quantum systems.

The manuscript is organized as follows: in Section~\ref{Sec:2}, we define the model Hamiltonian, the Keldysh field-theoretical framework that we employ, and the correlation function to be characterized quantitatively in the remainder of the manuscript. In Section~\ref{Sec:3}, we present the main analytical result of the present work, i.e., the linked-cluster approximation for the finite-temperature single-particle Green's function, thereby considering distinct momentum regimes and the associated long-time properties of the aforementioned single-particle Green's function. Then, in Section~\ref{Sec:4}, the obtained results are discussed in their physical content through the definition of a characteristic time-scale for the temperature-induced degradation of the zero-temperature results, as well as through the connection with a microscopic model. Finally, in Section~\ref{Sec:5}, we summarize the main findings and discuss further theoretical setups which might be amenable to further investigation in future studies.

\section{Model and methods} \label{Sec:2}

In this section, we introduce the physical model that we are going to consider, as well as the physical quantities that we are going to compute and the technical tools employed to this end.
Specifically, we consider the model Hamiltonian $\hat H = \hat H_{imp}+\hat H_{TLL}+\hat H_{int}$; the first contribution denotes the Hamiltonian of the bosonic impurity:
\begin{equation}
\hat H_{imp} = \sum_k \epsilon_k \hat d^{\dag}_k \hat d_k,
\end{equation}
where the impurity dispersion relation satisfies $\epsilon_{-k}=\epsilon_k$, the main focus of our work being the natural choice of a quadratic law of the form $\epsilon_k = t_i k^2$. The second term of the system's Hamiltonian is the Hamiltonian of the bath, taken to be of the form of a Tomonaga-Luttinger liquid (TLL)~\cite{giamarchi_book_1d} theory:
\begin{equation}
\hat H_{TLL}=\frac{v_s}{2\pi}\int dx \left[K(\partial_x\hat \theta)^2+\frac{1}{K}\left(\partial_x\hat\phi\right)^2 \right],
\end{equation}
where the fields $\hat\phi(x)$, $\hat \theta(x)$ satisfy the commutation relation $[\hat \phi(x),\partial_{x'}\hat \theta(x')]=i\pi \delta(x-x')$, $v_s$ is the sound velocity of the TLL and $K$ denotes the Luttinger parameter; the third contribution models instead the density-density interaction term between the TLL and the bosonic impurity:
\begin{align}
\hat H_{int}= -\frac{g}{\pi} \int dx\,\, (\partial_x \hat \phi)\,\, \hat d^{\dag}(x) \hat d(x),
\end{align}
where we introduced the notation $\hat d(x)=\frac{1}{\sqrt{L}}\sum_k e^{ikx} \hat d_k$. We remark that the choice of the form of $\hat H_{int}$ relies on the assumption that the Luttinger parameter $K$ of the 1D bath satisfies $K>1$, so that it is possible to retain only the long-wavelength part of the density fluctuations, while safely neglecting the Fourier modes at momenta that are multiples of $2\pi\rho_0$~\cite{Zvonarev_2007_Spin,Horovitz_2013_Collective}, $\rho_0$ being the number density of the 1D bath.

In order to provide a quantitative description of the dynamical properties of a mobile impurity inside a generic $1$D gapless quantum bath described by Luttinger-liquid theory, we initialize the system in the state:
\begin{equation}
    \hat\rho_0 = \frac{e^{-\beta\sum_{q\neq 0}v_s|q| \hat b^{\dag}_q \hat b_q}}{Z_{TLL}}\otimes \ket{0}\bra{0},
\end{equation}
where we introduced the notation $\ket{0}$ to denote the vacuum state for the impurity, and the bosonic operators $\hat b_q$ and $\hat b^{\dag}_q$ to indicate the bosonic modes that diagonalize $\hat H_{TLL}$, while $Z_{TLL}$ is the partition function of the thermal density matrix of the 1D bath. 

In the following, we evaluate the single-particle Green's function:
\begin{equation} \label{Eq:GF}
 G(y,t;x,0)=Tr\left[e^{i\hat H t} \hat d(y) e^{-i\hat H t} \hat d^{\dag}(x)\, \hat\rho_0 \right],
\end{equation}
in order to gain information on how the impurity's dynamics is affected by the interaction with the host 1D quantum liquid.

The technical framework that we rely on to perform the calculation of time-dependent correlation functions and observables is the Keldysh formalism~\cite{Kamenev2011}, where one introduces the Keldysh action $S[\psi,\psi^*]$ in order to express the Keldysh partition function as $Z=Tr[\hat\rho(t)]=\int \mathcal{D}[\psi,\psi^*] e^{iS[\psi,\psi^*]}=1$ as a function of the retarded and advanced fields $\psi=(\phi_{q,\pm},\varphi_{k,\pm})$ associated with the degrees of freedom of the system, to be defined below.
For our purposes, the action consists of $3$ terms, identified as the impurity action $S_{imp}$, the action of the TLL bath $S_{TLL}$ and the action $S_{int}$ describing the interaction between the latter two. In the following, we introduce in the different terms of the action the so called classical and quantum component fields $(\psi_{cl},\psi_{qu})$, respectively obtained from the retarded and advanced fields $\psi_{\pm}$ with identical set of quantum numbers via the rotation $\psi_{cl}=(\psi_+ +\psi_-)/\sqrt{2}$, $\psi_{qu} = (\psi_+-\psi_-)/\sqrt{2}$, where quantum number labels have been omitted for simplicity.

The Keldysh action for the system of noninteracting bosonic modes modelling the bath reads:
\begin{align}\label{Eq:S_TLL}
&S_{TLL}=\sum_{q\neq 0} \int \frac{d\omega}{2\pi} \sum_{\alpha,\beta = cl, qu} \phi^*_{q,\alpha}(\omega)\mathcal{G}^{-1}_{\alpha,\beta}\phi_{q,\beta}(\omega),\\
&\mathcal{G}^{-1}=\begin{pmatrix}
0 &  \omega-\omega_q-i\delta\\
 \omega-\omega_q+i\delta& 2i\delta F(\omega)
\end{pmatrix},
\end{align}
with $F(\omega)=1+2n_B(\omega)$ is simply related to the Bose-Einstein distribution $n_{B}(\omega)$ and $\omega_q=v_s|q|$.
The Keldysh action for the bosonic impurity has a similar structure:
\begin{align}\label{Eq:S_imp}
&S_{imp}=\sum_{k} \int \frac{d\epsilon}{2\pi} \sum_{\alpha,\beta = cl, qu} \varphi^*_{k,\alpha}(\epsilon)\mathcal{G}^{-1}_{\alpha,\beta}\varphi_{k,\beta}(\epsilon),\\
&\mathcal{G}^{-1}=\begin{pmatrix}
0 &  \epsilon-\epsilon_k-i\delta\\
 \epsilon-\epsilon_k+i\delta& 2i\delta F(\epsilon)
\end{pmatrix},
\end{align}
with $F(\epsilon)=1$ in the case of the initial density matrix being the vacuum state $\ket{0}\bra{0}$ for the impurity modes.
Finally, the interaction term takes the form:
\begin{equation}\label{Eq:S_int}
\begin{split}
S_{int} &=\int dt\,\, \left[-H_{int}\left(\phi_{+}(t),\varphi_+(t)\right)+H_{int}\left(\phi_{-}(t),\varphi_-(t)\right) \right] \\
&=-\frac{g\sqrt{K}}{L}\sum_k \sum_{p\neq 0}V(p)\int_{-\infty}^{+\infty} dt\\
&\Biggl\{\varphi^*_{k,+}(t) \varphi_{k+p,+}(t)\left[\phi^*_{p,+}(t)+\phi_{-p,+}(t) \right]+\\
&-\varphi^*_{k,-}(t) \varphi_{k+p,-}(t)\left[\phi^*_{p,-}(t)+\phi_{-p,-}(t) \right] \Biggr\}, 
\end{split}
\end{equation}
where $V(p)=\sqrt{\frac{L|p|}{2\pi}}e^{-\alpha\frac{|p|}{2}}$.

\section{Finite-temperature Green's function} \label{Sec:3}

In order to compute the single-particle Green's function \eqref{Eq:GF}, we reexpress it in the Keldysh formalism as:
\begin{multline}
G(y, t; x, 0) = 
\frac{1}{L}\sum_{k_1,k_2} e^{i(k_2 y -k_1 x)}\langle \varphi_{k_2,+}(t) \varphi^{*}_{k_1,+}(0) \rangle,
\end{multline}
where the average is taken over the action obtained by summing \eqref{Eq:S_TLL}, \eqref{Eq:S_imp}, \eqref{Eq:S_int}, together with the vacuum condition $F(\epsilon)=1$ in $S_{imp}$, while the bath is kept at finite temperature $T$. A second-order perturbative expansion in the interaction $S_{int}$ followed by a reexponentiation of the result, in the spirit of the linked-cluster expansion\cite{mahan_book}, leads to the final expression (see App.~\ref{SM:sec:GF_finite_T_LCE}):
\begin{equation}\label{Eq:GF_T}
\begin{split}
    &G(y,t;x,0)\approx \\
 &\approx\int \frac{dk}{2\pi}e^{ik(y-x)-i\epsilon_k t-\int du [R^{(a)}(u;k,T)+R^{(e)}(u;k,T)]\frac{1-iut-e^{-iut}}{u^2}},
 \end{split}
\end{equation}
where we introduced the notation:
\begin{widetext}
\begin{equation}
    \begin{split}
&R^{(a)}(u;k,T)=\frac{g^2 K}{(2\pi)^2}\int dp\,\, |p| e^{-\alpha|p|}n_B(\omega_p;T)\,\,\delta(u-\epsilon_{k+p}+\omega_p+\epsilon_k),\label{Eq:Ra}\\
&R^{(e)}(u;k,T)=\frac{g^2 K}{(2\pi)^2}\int dp\,\, |p| e^{-\alpha|p|}\Bigl(n_B(\omega_p;T)+1\Bigr)\,\,\delta(u-\epsilon_{k+p}-\omega_p+\epsilon_k),
\end{split}
\end{equation}
\end{widetext}
The interaction-induced correction term $R^{(e)}(u;k,T)$ accounts for the impurity's phonon emission processes, and is present even at zero temperature, while $R^{(a)}(u;k,T)$ is a genuine finite-temperature effect, capturing the impurity's phonon absorption processes. 

In what follows, we discuss how the finite bath temperature corrects the zero-temperature behavior of the momentum-space Green's function in the case of a quadratic dispersion relation $\epsilon_k = t_i k^2$; the latter is given by the expression:
\begin{equation}\label{Eq:momentum_GF}
    G(k,t;\beta)=e^{-i\epsilon_k t-\int du [R^{(a)}(u;k,T)+R^{(e)}(u;k,T)]\frac{1-iut-e^{-iut}}{u^2}},
\end{equation}
$\beta=1/T$ being the inverse temperature. At zero temperature, one has that $R^{(a)}(u;k,T)=0$, so that $G(k,t;\beta)$ satisfies:
\begin{equation}
G(k,t;\beta=+\infty)=e^{-i\epsilon_k t-\int du R^{(e)}(u;k,T=0)\frac{1-iut-e^{-iut}}{u^2}};
\end{equation}
the latter quantity has been carefully examined in Ref.~\cite{Kantian_2014_regimes} and is known to display a transition in its time dependence at the critical momentum $k_c = v_s/(2t_i)$, separating a standard quasiparticle behavior for $k> k_c$, associated with an exponential decay of $|G(k,t;\beta=+\infty)|$ as a function of time $t$, from an infrared-dominated behavior for $k<k_c$, signaled by a power-law decay of $|G(k,t;\beta=+\infty)|$ in time $t$. We provide a detailed account of the aforesaid decays in App.~\ref{SM:sec:zero_T}. 

We study the finite-temperature corrections by evaluating the ratio:
\begin{widetext}
\begin{equation}\label{Eq:ratio}
\begin{split}
  & \Biggl| \frac{G(k,t;\beta)}{G(k,t;+\infty)}\Biggr|=\\
   &=\exp\left\{-\frac{g^2 K}{(2\pi)^2}\int dp |p| e^{-\alpha |p|} n_B(\omega_p) \left[\frac{1-\cos[(t_i(k+p)^2-t_i k^2-v_s |p|)t]}{(t_i(k+p)^2-t_i k^2-v_s |p|)^2}+\frac{1-\cos[(t_i(k+p)^2-t_i k^2+v_s |p|)t]}{(t_i(k+p)^2-t_i k^2+v_s |p|)^2} \right]\right\},
   \end{split}
\end{equation}
\end{widetext}
and estimating its large-time behavior (see App.~\ref{S:G_k_large_t} for details). 
To this end, we observe that the dominant contributions come from the momentum values that fulfill the resonance condition $t_i(k+p)^2-t_i k^2 -v_s|p|=0$ for the phonon absorption processes and the resonance condition $t_i(k+p)^2-t_i k^2 + v_s|p|=0$ for the phonon emission processes. As detailed in App.~\ref{S:G_k_large_t}, we combine the knowledge of the poles of the aforementioned resonance conditions with the approximation of the delta function $\delta(x) = \lim_{t\rightarrow +\infty }\frac{1-\cos(tx)}{\pi t x^2}$, to obtain large-time estimates of the behavior of the momentum-space Green's function.  

\subsection{The IR regime $k<v_s/(2t_i)$}

%Let us start by considering the case $k=0$. In this case, Eq.\eqref{Eq:ratio} is considered for $k=0$, where one can check that the first term of the argument of the integral vanishes at $p=0$ and at a pair of opposite-sign momenta $p=\pm v_s/t_i$, while the second term has a pole at $p=0$ only. The explicit analytical evaluation of the large-time asymptotic behavior of the ration in Eq.\eqref{Eq:ratio} results in the following expression:
%\begin{equation} \label{Eq:k_0}
%\begin{split}
 %   &\Biggl| \frac{G(k,t;\beta)}{G(k,t;+\infty)}\Biggr|\approx\exp \Bigl\{ -\frac{g^2 K}{(2\pi)^2} \left[ \frac{2\pi t}{\beta v_s^2}+\frac{2\pi t e^{-\alpha v_s/t_i}}{t_i (e^{\beta v_s^2 /t_i}-1)} \right] \Bigr\}.
%    \end{split}
%\end{equation}
%for $t\gg 1$. As Eq.\eqref{Eq:k_0} manifestly shows, the effect of a finite-temperature consists in the appearance of a new set of zeros of the energy resonance conditions, namely the ones stemming from the first term in Eq.\eqref{Eq:ratio}, as well as the linear-in-time behavior of the corresponding contribution to the integral of Eq.\eqref{Eq:ratio} that the zero is associated to. The net effect is an exponential suppression of the zero-temperature result in time, whose characteristic timescale decreases as temperature increases. 

We start our analysis by characterizing the IR regime $0\leq k<v_s/(2t_i)$, thereby characterizing the large-time asymptotics of the finite-temperature correction to the single-particle Green's function. The zeros of the resonance condition for the phonon absorption process occur now at $p=0$, $p=(v_s -2 t_i k)/t_i$ and $p=(v_s +2 t_i k)/t_i$, while the phonon emission process is resonant only for $p=0$. The large-time asymptotic analysis, treated in detail in App.~\ref{S:G_k_large_t}, results ultimately in the following exponential suppression of the zero-temperature result:
\begin{equation} \label{Eq:k<}
  \begin{split}
    &\Biggl| \frac{G(k,t;\beta)}{G(k,t;+\infty)}\Biggr|\approx\\
    &\approx\exp \Biggl\{ -\frac{g^2 K}{(2\pi)^2}\pi t \Biggl[\frac{1}{\beta v_s}\left(\frac{1}{v_s - 2 t_i k}+ \frac{1}{v_s + 2 t_i k}\right)+\\
    &+\frac{e^{-\alpha(v_s -2t_i k)/t_i}}{t_i\left(e^{\beta v_s (v_s -2t_i k)/t_i} -1 \right)} + \frac{e^{-\alpha(v_s +2t_i k)/t_i}}{t_i\left(e^{\beta v_s (v_s +2t_i k)/t_i} -1 \right)} \Biggr] \Biggr\},
    \end{split}  
\end{equation}
where the precise meaning of the asymptotic estimate shown in Eq.~\eqref{Eq:k<}, as well as in the remainder of the main text, is discussed in App.~\ref{S:convergence}. As Eq.\eqref{Eq:k<} manifestly shows, the effect of a finite-temperature consists in the appearance of a new set of zeros of the energy resonance conditions, namely the ones stemming from the first term in Eq.\eqref{Eq:ratio}; the correction to the zero-temperature result displays moreover a linear-in-time behavior of the leading-order contribution to the integrals of Eq.\eqref{Eq:ratio} in the large-time limit. The net effect is an exponential suppression of the zero-temperature result in time, whose characteristic timescale decreases as temperature increases.

\begin{figure}[t]
\includegraphics[width=0.5\textwidth]{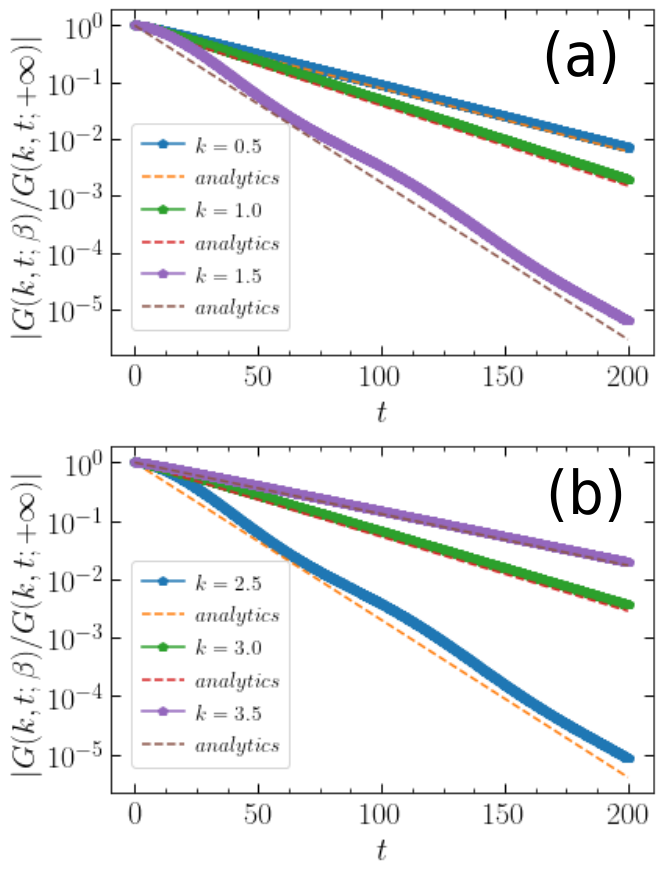}
\caption{(a) Ratio $|G(k,t;\beta)|/|G(k,t;+\infty)|$ as a function of time $t$ for parameter values $\beta = 2.0,\,
g = 0.5,\,
K = 1.2,\,
t_i = 0.25,\,
v_s = 1.0,\,
\alpha = 0.5$  with the choices of momentum $k=0.5,\, 1.0,\, 1.5$. (b) Ratio $|G(k,t;\beta)|/|G(k,t;+\infty)|$ as a function of time $t$ for parameter values $\beta = 2.0,\,
g = 0.5,\,
K = 1.2,\,
t_i = 0.25,\,
v_s = 1.0,\,
\alpha = 0.5$  with the choices of momentum $k=2.5,\, 3.0,\, 3.5$.}
\label{Fig:GF_k_0}
\end{figure}

In Fig.~\ref{Fig:GF_k_0}(a), we present some examples of the behavior of the ratio $|G(k,t;\beta)|/|G(k,t;+\infty)|$ given by Eq.~\eqref{Eq:ratio}, showing that it displays a trend that is compatible with the predicted large-time exponential decay predicted by Eq.~\eqref{Eq:k<}.

\subsection{The quasiparticle regime $k>v_s/(2t_i)$}

Moving on to the case where the momentum $k$ satisfies $k>v_s/(2t_i)$, we are able to show that the resonance condition for the phonon absorption processes is satisfied for momenta $p=0$ and $p=(v_s + 2t_i k)/t_i$, while the phonon emission processes are resonant at $p=0$ and $p=(2t_i k-v_s)/t_i$. The final result for the ratio of the finite-temperature and zero-temperature Green's functions is qualitatively analogous to the one found in the IR regime, despite the change in the distribution of the zeros of the resonance conditions. As shown in App.~\ref{S:G_k_large_t}, the explicit formula takes the form:
\begin{equation}\label{Eq:k>}
  \begin{split}
    &\Biggl| \frac{G(k,t;\beta)}{G(k,t;+\infty)}\Biggr|\approx\\
    &\approx\exp \Biggl\{ -\frac{g^2 K}{(2\pi)^2}\pi t \Biggl[\frac{1}{\beta v_s}\left(\frac{1}{2 t_i k - v_s}+ \frac{1}{v_s + 2 t_i k}\right)+\\
    &+\frac{e^{-\alpha(2t_i k-v_s)/t_i}}{t_i\left(e^{\beta v_s (2t_i k-v_s)/t_i} -1 \right)} + \frac{e^{-\alpha(v_s +2t_i k)/t_i}}{t_i\left(e^{\beta v_s (v_s +2t_i k)/t_i} -1 \right)} \Biggr] \Biggr\}.
    \end{split}  
\end{equation}
In the same spirit as for the IR regime, we present in Fig.~\ref{Fig:GF_k_0}(b) a few instances of the behavior of the ratio $|G(k,t;\beta)|/|G(k,t;+\infty)|$ given by Eq.~\eqref{Eq:ratio}, thereby providing evidence that it displays a trend that is compatible with the predicted large-time exponential decay predicted by Eq.~\eqref{Eq:k>}.

\begin{figure}[t]
\includegraphics[width=0.5\textwidth]{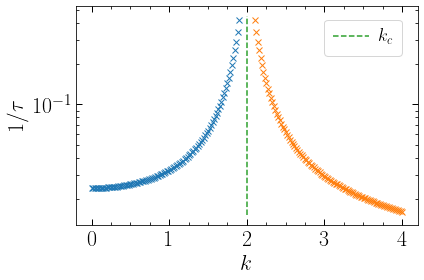}
\caption{Behavior of the inverse coherence time $1/\tau$ as a function of momentum $k$ for parameter values $
\beta = 2.0,\,
g = 0.5,\,
K = 1.2,\,
t_i = 0.25,\,
v_s = 1.0,\,
\alpha = 0.5$. The dashed green line highlights the location of the momentum $k_c=v_s/(2t_i)$ on the $x$-axis.}
\label{Fig:tau}
\end{figure}

\subsection{The special momentum value $k=v_s/(2t_i)$}

Finally, it is worth remarking that, for $k=k_c$, the finite-temperature correction to the momentum-space Green's function takes a different form from the ones already presented. Indeed, the result that we find for the finite-temperature correction to the momentum-space Green function takes the leading-order form: 
\begin{equation}\label{Eq:finite_T_corr_kc}
\begin{split}
    \frac{|G(k,t;\beta)|}{|G(k,t;+\infty)|} \approx\exp \Biggl\{-\frac{g^2 K}{(2\pi)^2} \left[\frac{2}{3\beta v_s} \sqrt{\frac{2\pi}{t_i}} t^{3/2}  \right]\Biggr\},
    \end{split}
\end{equation}
up to subleading linear-in-time corrections in the argument of the exponent (see App.~\ref{S:G_k_large_t}). Such an exceptional feature is responsible for the divergence of the finite-temperature coherence time $\tau$, to be introduced in Sec.~\ref{Sec:4}, as $k\rightarrow k_c^{\pm}$ (see Fig.~\ref{Fig:tau}).

\section{Discussion of the results} \label{Sec:4}

The leading behavior of the finite-temperature correction can be cast in the form:
\begin{equation}
    \Biggl| \frac{G(k,t;\beta)}{G(k,t;+\infty)}\Biggr|\approx \exp\left\{-t/\tau(k;g,v_s,K,t_i,T)\right\},
\end{equation}
where the expression of the inverse lifetime can be read off from Eqs.\eqref{Eq:k<}-\eqref{Eq:k>}.
Thus, the effect of temperature can be interpreted through the introduction of a finite coherence time beyond which the zero-temperature properties of the impurity dynamics within the bath are fundamentally altered as a result of the presence of a finite temperature.

We start the discussion by plotting the inverse lifetime $1/\tau$ as a function of momentum $k$ in Fig.~\ref{Fig:tau} while fixing the other parameter values. We underline two features of the displayed curve, namely the divergence at $k=k_c$ and the emergent symmetry between the blue and orange curves in the neighborhood of $k_c$. The divergence is readily explained in light of the non-analytic $t^{3/2}$ time-dependence in the leading behavior of the finite-temperature correction in Eq.~\eqref{Eq:ratio} when $k=k_c$ (see Eq.~\eqref{Eq:finite_T_corr_kc}), which formally causes $\tau^{-1}$ to diverge as $t\rightarrow +\infty$; on the other hand, the emergent symmetry of the behavior of $1/\tau$ in the left and right neighborhoods of $k_c$ traces back to the fact that the leading terms in the function $1/\tau$ as $k\rightarrow k_c^{\pm}$ depend on the distance $|k-k_c|$, so that $1/\tau$ becomes to leading order a function of $|k-k_c|$ only as $k\rightarrow k_c^+$ and $k\rightarrow k_c^-$, respectively. 

A further noteworthy observation is the effect of the impurity-bath interaction on the coherence time $\tau$: one has that $\tau\propto g^{-2}$, as expected, since a larger value of $g$ is associated with a more effective hybridization of the impurity with the bath degrees of freedom.

In order to gain further physical insight into the dependence of $\tau^{-1}$ on the microscopic parameters, we discuss the instance where the Luttinger liquid theory considered in the present work is an effective description of the low-energy excitations of the Lieb-Liniger model~\cite{Cazalilla_2011_bosons}. Since such model is Bethe-Ansatz solvable~\cite{Lieb_1963_liniger}, we are able to relate the phenomenological parameters $v_s$ and $K$ to the microscopic bath-bath interactions. Within such model, one defines the dimensionless parameter $\gamma=mg_{int}/\rho_0$, where $m$ is the bosons' mass, $g_{int}$ the strength of their contact interaction and $\rho_0$ their number density. Then, the curve $K(\gamma)$ can be plotted and analytically estimated in the small and large $\gamma$ limits, while Galilean invariance of the effective Luttinger liquid field theory allows to write the sound velocity as $v_s (\gamma) =\pi \rho_0/\left(m K(\gamma)\right)$. 

In the strong-coupling regime, one has, to leading order, that $K(\gamma)\approx 1$ and $v_s(\gamma)\approx \pi \rho_0/m := v_F$ (where the notation $v_F$ underlines the fact that the combination $\pi \rho_0/m$ equals the Fermi velocity of a free electron gas of density $\rho_0$ and mass $m$)~\cite{Cazalilla_2011_bosons}, so that the inverse coherence time becomes independent of the intra-bath interaction strength $g_{int}$ and converges to the value:
\begin{widetext}
\begin{equation}
   \frac{1}{\tau}=  \frac{g^2 }{4\pi}  \Biggl[\frac{1}{\beta v_F}\left(\frac{1}{v_F - 2 t_i k}+ \frac{1}{v_F + 2 t_i k}\right)
    +\frac{e^{-\alpha(v_F -2t_i k)/t_i}}{t_i\left(e^{\beta v_F (v_F -2t_i k)/t_i} -1 \right)} + \frac{e^{-\alpha(v_F +2t_i k)/t_i}}{t_i\left(e^{\beta v_F (v_F +2t_i k)/t_i} -1 \right)} \Biggr] 
\end{equation}
for $k<v_s/(2t_i)=\pi \rho_0/(2mt_i)$, and to the value:
\begin{equation}
    \frac{1}{\tau}= \frac{g^2 }{4\pi} \Biggl[\frac{1}{\beta v_F}\left(\frac{1}{2 t_i k - v_F}+ \frac{1}{v_F + 2 t_i k}\right)
    +\frac{e^{-\alpha(2t_i k-v_F)/t_i}}{t_i\left(e^{\beta v_F (2t_i k-v_F)/t_i} -1 \right)} + \frac{e^{-\alpha(v_F +2t_i k)/t_i}}{t_i\left(e^{\beta v_F (v_F +2t_i k)/t_i} -1 \right)} \Biggr]
\end{equation}
\end{widetext}
for the opposite scenario $k>v_s/(2t_i)=\pi \rho_0/(2mt_i)$.

In the weak-coupling regime, instead, one has that $K(\gamma)\approx \pi/\sqrt{\gamma}$, which results in $v_s(\gamma)\approx \rho_0 \sqrt{\gamma}/m$~\cite{Cazalilla_2011_bosons}. In order to estimate the inverse coherence time, we consider separately the IR and quasiparticle regimes. When $k<v_s/(2t_i)$, we provide an estimate of $\tau^{-1}$ in the limit $k\ll v_s(\gamma)/(2t_i)=\rho_0 \sqrt{\gamma}/(2m t_i) \ll 1$ by setting $k=0$ to leading order in Eq.~\eqref{Eq:k<} and considering the leading term in the limit of small $\gamma$, so that the inverse coherence time reads in this limit:
\begin{equation}
    \frac{1}{\tau} \approx \frac{g^2 K m^2}{\pi \beta  \rho_0^2 \gamma}.
\end{equation}
Thus, one observes that the temperature-induced inverse coherence time decreases, to a first approximation, as $\gamma^{-1}$ with the intrabath interaction strength $\gamma$.
On the other hand, when $k> v_s(\gamma)/(2t_i)=\rho_0 \sqrt{\gamma}/(2m t_i)$, the leading contribution to Eq.~\eqref{Eq:k>} in the limit of small values of $\gamma$ gives the result:
\begin{equation}
     \frac{1}{\tau} \approx \frac{g^2 K m (1+e^{-2\alpha k})}{4\pi \beta  t_i k \rho_0 \sqrt{\gamma}}.
\end{equation}
In such regime, we observe instead a $\gamma^{-1/2}$ dependence of the inverse coherence time on the intrabath interaction strength $\gamma$, differently from what predicted in the IR regime. Nevertheless, in both cases, in the limit $\gamma\ll 1$, the decrease in the strength of the interactions within the chosen Lieb-Liniger bath causes the temperature-induced coherence time to shrink. 

\section{Conclusion and outlook} \label{Sec:5}

In this work, we have investigated the effect of a finite temperature in the Luttinger liquid bath on the dynamical signatures of a mobile impurity that interacts with it through a density-density contact term. Within a linked-cluster approximation in the framework of a Keldysh field-theoretical representation of the system's dynamics, we have derived an analytical expression for the one-particle Green's function of the impurity in presence of a finite-temperature 1D bath. Focusing on the momentum-space Green's function, we unveiled that, regardless of the chosen parameter regime, the finite value of the temperature $T$ introduces a temperature- and momentum-dependent coherence time that sets the time-scale for coherent zero-temperature propagation of the impurity, and beyond which the aforesaid coherence is exponentially degraded.

The present results could be tested in experimental setups based on ultracold atoms, where the observation of impurity dynamics in 1D many-body systems is already within reach in state-of-the-art facilities~\cite{Catani_2012_bose,Fukuhara_2013_quantum,Meinert_2017_bloch,Horvath_2026_observing_impurity}. Moreover, while quantum gas microscopes naturally allow for direct access to the atoms' density profiles~\cite{Bakr_2009_microscope,Bakr_2010_probing,Sherson_2010_single,Haller_2015_single,Cheuk_2015_fermions,Gross_2021_detection}, there exist concrete theoretical proposals aiming at the measurement of single-particle Green's functions in strongly-correlated quantum gases~\cite{Knap_2013_ramsey,Jen_2013_eit,Jen_2014_extracting,Weitenberg_2026_protocols}. We leave the discussion of the interplay of such techniques with our results as a promising direction of future research. 

On the theoretical side, this work invites further investigations of the complex dynamics of mobile impurities in one-dimensional many-body environments. More specifically, it would be timely to discuss the role of the bath-impurity interaction range in the impurity's dynamical properties, by replacing the standard contact interaction with a structured finite-range or, potentially, long-range one. Moreover, 1D quantum many-body systems may realize several instabilities of the Luttinger-liquid theory; the exploration of the features of impurity dynamics within different 1D phases of matter for the bath is therefore a potential avenue for future detailed investigations.     

\textit{Acknowledgements}: This work was supported by the
Swiss National Science Foundation under grant number
200020-219400.

\bibliography{finite_temperature,totphys-A-J,totphys-K-Z}

%apsrev4-2.bst 2019-01-14 (MD) hand-edited version of apsrev4-1.bst
%Control: key (0)
%Control: author (8) initials jnrlst
%Control: editor formatted (1) identically to author
%Control: production of article title (0) allowed
%Control: page (0) single
%Control: year (1) truncated
%Control: production of eprint (0) enabled
\begin{thebibliography}{53}%
\makeatletter
\providecommand \@ifxundefined [1]{%
 \@ifx{#1\undefined}
}%
\providecommand \@ifnum [1]{%
 \ifnum #1\expandafter \@firstoftwo
 \else \expandafter \@secondoftwo
 \fi
}%
\providecommand \@ifx [1]{%
 \ifx #1\expandafter \@firstoftwo
 \else \expandafter \@secondoftwo
 \fi
}%
\providecommand \natexlab [1]{#1}%
\providecommand \enquote  [1]{``#1''}%
\providecommand \bibnamefont  [1]{#1}%
\providecommand \bibfnamefont [1]{#1}%
\providecommand \citenamefont [1]{#1}%
\providecommand \href@noop [0]{\@secondoftwo}%
\providecommand \href [0]{\begingroup \@sanitize@url \@href}%
\providecommand \@href[1]{\@@startlink{#1}\@@href}%
\providecommand \@@href[1]{\endgroup#1\@@endlink}%
\providecommand \@sanitize@url [0]{\catcode `\\12\catcode `\$12\catcode `\&12\catcode `\#12\catcode `\^12\catcode `\_12\catcode `\%12\relax}%
\providecommand \@@startlink[1]{}%
\providecommand \@@endlink[0]{}%
\providecommand \url  [0]{\begingroup\@sanitize@url \@url }%
\providecommand \@url [1]{\endgroup\@href {#1}{\urlprefix }}%
\providecommand \urlprefix  [0]{URL }%
\providecommand \Eprint [0]{\href }%
\providecommand \doibase [0]{https://doi.org/}%
\providecommand \selectlanguage [0]{\@gobble}%
\providecommand \bibinfo  [0]{\@secondoftwo}%
\providecommand \bibfield  [0]{\@secondoftwo}%
\providecommand \translation [1]{[#1]}%
\providecommand \BibitemOpen [0]{}%
\providecommand \bibitemStop [0]{}%
\providecommand \bibitemNoStop [0]{.\EOS\space}%
\providecommand \EOS [0]{\spacefactor3000\relax}%
\providecommand \BibitemShut  [1]{\csname bibitem#1\endcsname}%
\let\auto@bib@innerbib\@empty
%</preamble>
\bibitem [{\citenamefont {Zvonarev}\ \emph {et~al.}(2007)\citenamefont {Zvonarev}, \citenamefont {Cheianov},\ and\ \citenamefont {Giamarchi}}]{Zvonarev_2007_Spin}%
  \BibitemOpen
  \bibfield  {author} {\bibinfo {author} {\bibfnamefont {M.~B.}\ \bibnamefont {Zvonarev}}, \bibinfo {author} {\bibfnamefont {V.~V.}\ \bibnamefont {Cheianov}},\ and\ \bibinfo {author} {\bibfnamefont {T.}~\bibnamefont {Giamarchi}},\ }\bibfield  {title} {\bibinfo {title} {Spin dynamics in a one-dimensional ferromagnetic bose gas},\ }\href {https://doi.org/10.1103/PhysRevLett.99.240404} {\bibfield  {journal} {\bibinfo  {journal} {Phys. Rev. Lett.}\ }\textbf {\bibinfo {volume} {99}},\ \bibinfo {pages} {240404} (\bibinfo {year} {2007})}\BibitemShut {NoStop}%
\bibitem [{\citenamefont {Kantian}\ \emph {et~al.}(2014)\citenamefont {Kantian}, \citenamefont {Schollw\"ock},\ and\ \citenamefont {Giamarchi}}]{Kantian_2014_regimes}%
  \BibitemOpen
  \bibfield  {author} {\bibinfo {author} {\bibfnamefont {A.}~\bibnamefont {Kantian}}, \bibinfo {author} {\bibfnamefont {U.}~\bibnamefont {Schollw\"ock}},\ and\ \bibinfo {author} {\bibfnamefont {T.}~\bibnamefont {Giamarchi}},\ }\bibfield  {title} {\bibinfo {title} {Competing regimes of motion of 1d mobile impurities},\ }\href {https://doi.org/10.1103/PhysRevLett.113.070601} {\bibfield  {journal} {\bibinfo  {journal} {Phys. Rev. Lett.}\ }\textbf {\bibinfo {volume} {113}},\ \bibinfo {pages} {070601} (\bibinfo {year} {2014})}\BibitemShut {NoStop}%
\bibitem [{\citenamefont {Kamar}\ \emph {et~al.}(2025)\citenamefont {Kamar}, \citenamefont {Kantian},\ and\ \citenamefont {Giamarchi}}]{Kamar_2025_ladder}%
  \BibitemOpen
  \bibfield  {author} {\bibinfo {author} {\bibfnamefont {N.~A.}\ \bibnamefont {Kamar}}, \bibinfo {author} {\bibfnamefont {A.}~\bibnamefont {Kantian}},\ and\ \bibinfo {author} {\bibfnamefont {T.}~\bibnamefont {Giamarchi}},\ }\bibfield  {title} {\bibinfo {title} {Mobile impurity in a two-leg bosonic ladder},\ }\href {https://doi.org/10.1103/vh99-hdhh} {\bibfield  {journal} {\bibinfo  {journal} {Phys. Rev. Res.}\ }\textbf {\bibinfo {volume} {7}},\ \bibinfo {pages} {033136} (\bibinfo {year} {2025})}\BibitemShut {NoStop}%
\bibitem [{\citenamefont {Gotta}\ and\ \citenamefont {Giamarchi}(2024)}]{Gotta_2024_Dirac}%
  \BibitemOpen
  \bibfield  {author} {\bibinfo {author} {\bibfnamefont {L.}~\bibnamefont {Gotta}}\ and\ \bibinfo {author} {\bibfnamefont {T.}~\bibnamefont {Giamarchi}},\ }\bibfield  {title} {\bibinfo {title} {Dirac impurity in a luttinger liquid},\ }\href {https://doi.org/10.1103/PhysRevB.110.165116} {\bibfield  {journal} {\bibinfo  {journal} {Phys. Rev. B}\ }\textbf {\bibinfo {volume} {110}},\ \bibinfo {pages} {165116} (\bibinfo {year} {2024})}\BibitemShut {NoStop}%
\bibitem [{\citenamefont {Lamacraft}(2009)}]{Lamacraft_2009_spectral}%
  \BibitemOpen
  \bibfield  {author} {\bibinfo {author} {\bibfnamefont {A.}~\bibnamefont {Lamacraft}},\ }\bibfield  {title} {\bibinfo {title} {Dispersion relation and spectral function of an impurity in a one-dimensional quantum liquid},\ }\href {https://doi.org/10.1103/PhysRevB.79.241105} {\bibfield  {journal} {\bibinfo  {journal} {Phys. Rev. B}\ }\textbf {\bibinfo {volume} {79}},\ \bibinfo {pages} {241105(R)} (\bibinfo {year} {2009})}\BibitemShut {NoStop}%
\bibitem [{\citenamefont {Schecter}\ \emph {et~al.}(2012)\citenamefont {Schecter}, \citenamefont {Kamenev}, \citenamefont {Gangardt},\ and\ \citenamefont {Lamacraft}}]{Schecter_2012_velocity}%
  \BibitemOpen
  \bibfield  {author} {\bibinfo {author} {\bibfnamefont {M.}~\bibnamefont {Schecter}}, \bibinfo {author} {\bibfnamefont {A.}~\bibnamefont {Kamenev}}, \bibinfo {author} {\bibfnamefont {D.~M.}\ \bibnamefont {Gangardt}},\ and\ \bibinfo {author} {\bibfnamefont {A.}~\bibnamefont {Lamacraft}},\ }\bibfield  {title} {\bibinfo {title} {Critical velocity of a mobile impurity in one-dimensional quantum liquids},\ }\href {https://doi.org/10.1103/PhysRevLett.108.207001} {\bibfield  {journal} {\bibinfo  {journal} {Phys. Rev. Lett.}\ }\textbf {\bibinfo {volume} {108}},\ \bibinfo {pages} {207001} (\bibinfo {year} {2012})}\BibitemShut {NoStop}%
\bibitem [{\citenamefont {Burovski}\ \emph {et~al.}(2014)\citenamefont {Burovski}, \citenamefont {Cheianov}, \citenamefont {Gamayun},\ and\ \citenamefont {Lychkovskiy}}]{Burovski_2014_relaxation}%
  \BibitemOpen
  \bibfield  {author} {\bibinfo {author} {\bibfnamefont {E.}~\bibnamefont {Burovski}}, \bibinfo {author} {\bibfnamefont {V.}~\bibnamefont {Cheianov}}, \bibinfo {author} {\bibfnamefont {O.}~\bibnamefont {Gamayun}},\ and\ \bibinfo {author} {\bibfnamefont {O.}~\bibnamefont {Lychkovskiy}},\ }\bibfield  {title} {\bibinfo {title} {Momentum relaxation of a mobile impurity in a one-dimensional quantum gas},\ }\href {https://doi.org/10.1103/PhysRevA.89.041601} {\bibfield  {journal} {\bibinfo  {journal} {Phys. Rev. A}\ }\textbf {\bibinfo {volume} {89}},\ \bibinfo {pages} {041601(R)} (\bibinfo {year} {2014})}\BibitemShut {NoStop}%
\bibitem [{\citenamefont {Shen}\ and\ \citenamefont {Fan}(2007)}]{Shen_2007_transport}%
  \BibitemOpen
  \bibfield  {author} {\bibinfo {author} {\bibfnamefont {J.-T.}\ \bibnamefont {Shen}}\ and\ \bibinfo {author} {\bibfnamefont {S.}~\bibnamefont {Fan}},\ }\bibfield  {title} {\bibinfo {title} {Strongly correlated multiparticle transport in one dimension through a quantum impurity},\ }\href {https://doi.org/10.1103/PhysRevA.76.062709} {\bibfield  {journal} {\bibinfo  {journal} {Phys. Rev. A}\ }\textbf {\bibinfo {volume} {76}},\ \bibinfo {pages} {062709} (\bibinfo {year} {2007})}\BibitemShut {NoStop}%
\bibitem [{\citenamefont {Enss}\ \emph {et~al.}(2005)\citenamefont {Enss}, \citenamefont {Meden}, \citenamefont {Andergassen}, \citenamefont {Barnab\'e-Th\'eriault}, \citenamefont {Metzner},\ and\ \citenamefont {Sch\"onhammer}}]{Enss_2005_correlation}%
  \BibitemOpen
  \bibfield  {author} {\bibinfo {author} {\bibfnamefont {T.}~\bibnamefont {Enss}}, \bibinfo {author} {\bibfnamefont {V.}~\bibnamefont {Meden}}, \bibinfo {author} {\bibfnamefont {S.}~\bibnamefont {Andergassen}}, \bibinfo {author} {\bibfnamefont {X.}~\bibnamefont {Barnab\'e-Th\'eriault}}, \bibinfo {author} {\bibfnamefont {W.}~\bibnamefont {Metzner}},\ and\ \bibinfo {author} {\bibfnamefont {K.}~\bibnamefont {Sch\"onhammer}},\ }\bibfield  {title} {\bibinfo {title} {Impurity and correlation effects on transport in one-dimensional quantum wires},\ }\href {https://doi.org/10.1103/PhysRevB.71.155401} {\bibfield  {journal} {\bibinfo  {journal} {Phys. Rev. B}\ }\textbf {\bibinfo {volume} {71}},\ \bibinfo {pages} {155401} (\bibinfo {year} {2005})}\BibitemShut {NoStop}%
\bibitem [{\citenamefont {Sacha}\ and\ \citenamefont {Timmermans}(2006)}]{Sacha_2006_bec}%
  \BibitemOpen
  \bibfield  {author} {\bibinfo {author} {\bibfnamefont {K.}~\bibnamefont {Sacha}}\ and\ \bibinfo {author} {\bibfnamefont {E.}~\bibnamefont {Timmermans}},\ }\bibfield  {title} {\bibinfo {title} {Self-localized impurities embedded in a one-dimensional bose-einstein condensate and their quantum fluctuations},\ }\href {https://doi.org/10.1103/PhysRevA.73.063604} {\bibfield  {journal} {\bibinfo  {journal} {Phys. Rev. A}\ }\textbf {\bibinfo {volume} {73}},\ \bibinfo {pages} {063604} (\bibinfo {year} {2006})}\BibitemShut {NoStop}%
\bibitem [{\citenamefont {Lesage}\ and\ \citenamefont {Saleur}(1997)}]{Lesage_1997_correlations}%
  \BibitemOpen
  \bibfield  {author} {\bibinfo {author} {\bibfnamefont {F.}~\bibnamefont {Lesage}}\ and\ \bibinfo {author} {\bibfnamefont {H.}~\bibnamefont {Saleur}},\ }\bibfield  {title} {\bibinfo {title} {Correlations in one-dimensional quantum impurity problems with an external field},\ }\href@noop {} {\bibfield  {journal} {\bibinfo  {journal} {Nuclear physics B}\ }\textbf {\bibinfo {volume} {490}},\ \bibinfo {pages} {543} (\bibinfo {year} {1997})}\BibitemShut {NoStop}%
\bibitem [{\citenamefont {Bonart}\ and\ \citenamefont {Cugliandolo}(2012)}]{Bonart_2012_brownian}%
  \BibitemOpen
  \bibfield  {author} {\bibinfo {author} {\bibfnamefont {J.}~\bibnamefont {Bonart}}\ and\ \bibinfo {author} {\bibfnamefont {L.~F.}\ \bibnamefont {Cugliandolo}},\ }\bibfield  {title} {\bibinfo {title} {From nonequilibrium quantum brownian motion to impurity dynamics in one-dimensional quantum liquids},\ }\href {https://doi.org/10.1103/PhysRevA.86.023636} {\bibfield  {journal} {\bibinfo  {journal} {Phys. Rev. A}\ }\textbf {\bibinfo {volume} {86}},\ \bibinfo {pages} {023636} (\bibinfo {year} {2012})}\BibitemShut {NoStop}%
\bibitem [{\citenamefont {Haldane}(1981)}]{Haldane_1981_Luttinger}%
  \BibitemOpen
  \bibfield  {author} {\bibinfo {author} {\bibfnamefont {F.~D.~M.}\ \bibnamefont {Haldane}},\ }\bibfield  {title} {\bibinfo {title} {'luttinger liquid theory'of one-dimensional quantum fluids. i. properties of the luttinger model and their extension to the general 1d interacting spinless fermi gas},\ }\href@noop {} {\bibfield  {journal} {\bibinfo  {journal} {Journal of Physics C: Solid State Physics}\ }\textbf {\bibinfo {volume} {14}},\ \bibinfo {pages} {2585} (\bibinfo {year} {1981})}\BibitemShut {NoStop}%
\bibitem [{\citenamefont {Giamarchi}(2003)}]{Giamarchi_2003_1D}%
  \BibitemOpen
  \bibfield  {author} {\bibinfo {author} {\bibfnamefont {T.}~\bibnamefont {Giamarchi}},\ }\href@noop {} {\emph {\bibinfo {title} {Quantum physics in one dimension}}},\ Vol.\ \bibinfo {volume} {121}\ (\bibinfo  {publisher} {Clarendon press},\ \bibinfo {year} {2003})\BibitemShut {NoStop}%
\bibitem [{\citenamefont {Giamarchi}(2004)}]{giamarchi_book_1d}%
  \BibitemOpen
  \bibfield  {author} {\bibinfo {author} {\bibfnamefont {T.}~\bibnamefont {Giamarchi}},\ }\href@noop {} {\emph {\bibinfo {title} {Quantum Physics in One Dimension}}},\ \bibinfo {series} {International series of monographs on physics}, Vol.\ \bibinfo {volume} {121}\ (\bibinfo  {publisher} {Oxford University Press},\ \bibinfo {address} {Oxford},\ \bibinfo {year} {2004})\BibitemShut {NoStop}%
\bibitem [{\citenamefont {Bouchoule}\ \emph {et~al.}(2025)\citenamefont {Bouchoule}, \citenamefont {Citro}, \citenamefont {Duty}, \citenamefont {Giamarchi}, \citenamefont {Hulet}, \citenamefont {Klanj{\v{s}}ek}, \citenamefont {Orignac},\ and\ \citenamefont {Weber}}]{Bouchoule_2025_Platforms}%
  \BibitemOpen
  \bibfield  {author} {\bibinfo {author} {\bibfnamefont {I.}~\bibnamefont {Bouchoule}}, \bibinfo {author} {\bibfnamefont {R.}~\bibnamefont {Citro}}, \bibinfo {author} {\bibfnamefont {T.}~\bibnamefont {Duty}}, \bibinfo {author} {\bibfnamefont {T.}~\bibnamefont {Giamarchi}}, \bibinfo {author} {\bibfnamefont {R.~G.}\ \bibnamefont {Hulet}}, \bibinfo {author} {\bibfnamefont {M.}~\bibnamefont {Klanj{\v{s}}ek}}, \bibinfo {author} {\bibfnamefont {E.}~\bibnamefont {Orignac}},\ and\ \bibinfo {author} {\bibfnamefont {B.}~\bibnamefont {Weber}},\ }\bibfield  {title} {\bibinfo {title} {Platforms for the realization and characterization of tomonaga--luttinger liquids},\ }\href@noop {} {\bibfield  {journal} {\bibinfo  {journal} {Nature Reviews Physics}\ }\textbf {\bibinfo {volume} {7}},\ \bibinfo {pages} {565} (\bibinfo {year} {2025})}\BibitemShut {NoStop}%
\bibitem [{\citenamefont {Sch{\"o}nhammer}(2004)}]{Schonhammer_2004_Luttinger}%
  \BibitemOpen
  \bibfield  {author} {\bibinfo {author} {\bibfnamefont {K.}~\bibnamefont {Sch{\"o}nhammer}},\ }\bibfield  {title} {\bibinfo {title} {Luttinger liquids: the basic concepts},\ }\href@noop {} {\bibfield  {journal} {\bibinfo  {journal} {Strong interactions in low dimensions}\ ,\ \bibinfo {pages} {93}} (\bibinfo {year} {2004})}\BibitemShut {NoStop}%
\bibitem [{\citenamefont {Fisher}\ and\ \citenamefont {Glazman}(1997)}]{Fisher_1997_transport}%
  \BibitemOpen
  \bibfield  {author} {\bibinfo {author} {\bibfnamefont {M.~P.}\ \bibnamefont {Fisher}}\ and\ \bibinfo {author} {\bibfnamefont {L.~I.}\ \bibnamefont {Glazman}},\ }\bibfield  {title} {\bibinfo {title} {Transport in a one-dimensional luttinger liquid},\ }in\ \href@noop {} {\emph {\bibinfo {booktitle} {Mesoscopic electron transport}}}\ (\bibinfo  {publisher} {Springer},\ \bibinfo {year} {1997})\ pp.\ \bibinfo {pages} {331--373}\BibitemShut {NoStop}%
\bibitem [{\citenamefont {Pham}\ \emph {et~al.}(2000)\citenamefont {Pham}, \citenamefont {Gabay},\ and\ \citenamefont {Lederer}}]{Pham_2000_fractional}%
  \BibitemOpen
  \bibfield  {author} {\bibinfo {author} {\bibfnamefont {K.-V.}\ \bibnamefont {Pham}}, \bibinfo {author} {\bibfnamefont {M.}~\bibnamefont {Gabay}},\ and\ \bibinfo {author} {\bibfnamefont {P.}~\bibnamefont {Lederer}},\ }\bibfield  {title} {\bibinfo {title} {Fractional excitations in the luttinger liquid},\ }\href {https://doi.org/10.1103/PhysRevB.61.16397} {\bibfield  {journal} {\bibinfo  {journal} {Phys. Rev. B}\ }\textbf {\bibinfo {volume} {61}},\ \bibinfo {pages} {16397} (\bibinfo {year} {2000})}\BibitemShut {NoStop}%
\bibitem [{\citenamefont {Castro~Neto}\ and\ \citenamefont {Fisher}(1996)}]{Neto_1996_heavy}%
  \BibitemOpen
  \bibfield  {author} {\bibinfo {author} {\bibfnamefont {A.~H.}\ \bibnamefont {Castro~Neto}}\ and\ \bibinfo {author} {\bibfnamefont {M.~P.~A.}\ \bibnamefont {Fisher}},\ }\bibfield  {title} {\bibinfo {title} {Dynamics of a heavy particle in a luttinger liquid},\ }\href {https://doi.org/10.1103/PhysRevB.53.9713} {\bibfield  {journal} {\bibinfo  {journal} {Phys. Rev. B}\ }\textbf {\bibinfo {volume} {53}},\ \bibinfo {pages} {9713} (\bibinfo {year} {1996})}\BibitemShut {NoStop}%
\bibitem [{\citenamefont {Cheianov}\ and\ \citenamefont {Zvonarev}(2004{\natexlab{a}})}]{Cheianov_2004_Nonunitary}%
  \BibitemOpen
  \bibfield  {author} {\bibinfo {author} {\bibfnamefont {V.~V.}\ \bibnamefont {Cheianov}}\ and\ \bibinfo {author} {\bibfnamefont {M.~B.}\ \bibnamefont {Zvonarev}},\ }\bibfield  {title} {\bibinfo {title} {Nonunitary spin-charge separation in a one-dimensional fermion gas},\ }\href {https://doi.org/10.1103/PhysRevLett.92.176401} {\bibfield  {journal} {\bibinfo  {journal} {Phys. Rev. Lett.}\ }\textbf {\bibinfo {volume} {92}},\ \bibinfo {pages} {176401} (\bibinfo {year} {2004}{\natexlab{a}})}\BibitemShut {NoStop}%
\bibitem [{\citenamefont {Cheianov}\ and\ \citenamefont {Zvonarev}(2004{\natexlab{b}})}]{Cheianov_2004_Zero}%
  \BibitemOpen
  \bibfield  {author} {\bibinfo {author} {\bibfnamefont {V.~V.}\ \bibnamefont {Cheianov}}\ and\ \bibinfo {author} {\bibfnamefont {M.}~\bibnamefont {Zvonarev}},\ }\bibfield  {title} {\bibinfo {title} {Zero temperature correlation functions for the impenetrable fermion gas},\ }\href@noop {} {\bibfield  {journal} {\bibinfo  {journal} {Journal of Physics A: Mathematical and General}\ }\textbf {\bibinfo {volume} {37}},\ \bibinfo {pages} {2261} (\bibinfo {year} {2004}{\natexlab{b}})}\BibitemShut {NoStop}%
\bibitem [{\citenamefont {Massel}\ \emph {et~al.}(2013)\citenamefont {Massel}, \citenamefont {Kantian}, \citenamefont {Daley}, \citenamefont {Giamarchi},\ and\ \citenamefont {T{\"o}rm{\"a}}}]{Massel_2013_dynamics}%
  \BibitemOpen
  \bibfield  {author} {\bibinfo {author} {\bibfnamefont {F.}~\bibnamefont {Massel}}, \bibinfo {author} {\bibfnamefont {A.}~\bibnamefont {Kantian}}, \bibinfo {author} {\bibfnamefont {A.~J.}\ \bibnamefont {Daley}}, \bibinfo {author} {\bibfnamefont {T.}~\bibnamefont {Giamarchi}},\ and\ \bibinfo {author} {\bibfnamefont {P.}~\bibnamefont {T{\"o}rm{\"a}}},\ }\bibfield  {title} {\bibinfo {title} {Dynamics of an impurity in a one-dimensional lattice},\ }\href@noop {} {\bibfield  {journal} {\bibinfo  {journal} {New Journal of Physics}\ }\textbf {\bibinfo {volume} {15}},\ \bibinfo {pages} {045018} (\bibinfo {year} {2013})}\BibitemShut {NoStop}%
\bibitem [{\citenamefont {Kamar}\ \emph {et~al.}(2019)\citenamefont {Kamar}, \citenamefont {Kantian},\ and\ \citenamefont {Giamarchi}}]{Kamar_2019_Dynamics}%
  \BibitemOpen
  \bibfield  {author} {\bibinfo {author} {\bibfnamefont {N.~A.}\ \bibnamefont {Kamar}}, \bibinfo {author} {\bibfnamefont {A.}~\bibnamefont {Kantian}},\ and\ \bibinfo {author} {\bibfnamefont {T.}~\bibnamefont {Giamarchi}},\ }\bibfield  {title} {\bibinfo {title} {Dynamics of a mobile impurity in a two-leg bosonic ladder},\ }\href {https://doi.org/10.1103/PhysRevA.100.023614} {\bibfield  {journal} {\bibinfo  {journal} {Phys. Rev. A}\ }\textbf {\bibinfo {volume} {100}},\ \bibinfo {pages} {023614} (\bibinfo {year} {2019})}\BibitemShut {NoStop}%
\bibitem [{\citenamefont {Anderson}(1967)}]{Anderson_1967_aoc}%
  \BibitemOpen
  \bibfield  {author} {\bibinfo {author} {\bibfnamefont {P.~W.}\ \bibnamefont {Anderson}},\ }\bibfield  {title} {\bibinfo {title} {Infrared catastrophe in fermi gases with local scattering potentials},\ }\href {https://doi.org/10.1103/PhysRevLett.18.1049} {\bibfield  {journal} {\bibinfo  {journal} {Phys. Rev. Lett.}\ }\textbf {\bibinfo {volume} {18}},\ \bibinfo {pages} {1049} (\bibinfo {year} {1967})}\BibitemShut {NoStop}%
\bibitem [{\citenamefont {Qin}\ \emph {et~al.}(1996)\citenamefont {Qin}, \citenamefont {Fabrizio},\ and\ \citenamefont {Yu}}]{Qin_1996_catastrophe}%
  \BibitemOpen
  \bibfield  {author} {\bibinfo {author} {\bibfnamefont {S.}~\bibnamefont {Qin}}, \bibinfo {author} {\bibfnamefont {M.}~\bibnamefont {Fabrizio}},\ and\ \bibinfo {author} {\bibfnamefont {L.}~\bibnamefont {Yu}},\ }\bibfield  {title} {\bibinfo {title} {Impurity in a luttinger liquid: A numerical study of the finite-size energy spectrum and of the orthogonality catastrophe exponent},\ }\href {https://doi.org/10.1103/PhysRevB.54.R9643} {\bibfield  {journal} {\bibinfo  {journal} {Phys. Rev. B}\ }\textbf {\bibinfo {volume} {54}},\ \bibinfo {pages} {R9643(R)} (\bibinfo {year} {1996})}\BibitemShut {NoStop}%
\bibitem [{\citenamefont {Furusaki}(1997)}]{Furusaki_1997_orthogonality}%
  \BibitemOpen
  \bibfield  {author} {\bibinfo {author} {\bibfnamefont {A.}~\bibnamefont {Furusaki}},\ }\bibfield  {title} {\bibinfo {title} {Local perturbation in a tomonaga-luttinger liquid at $g=1/2:$ orthogonality catastrophe, fermi-edge singularity, and local density of states},\ }\href {https://doi.org/10.1103/PhysRevB.56.9352} {\bibfield  {journal} {\bibinfo  {journal} {Phys. Rev. B}\ }\textbf {\bibinfo {volume} {56}},\ \bibinfo {pages} {9352} (\bibinfo {year} {1997})}\BibitemShut {NoStop}%
\bibitem [{\citenamefont {Caracanhas}\ and\ \citenamefont {Pereira}(2015)}]{Caracanhas_2015_Vortex}%
  \BibitemOpen
  \bibfield  {author} {\bibinfo {author} {\bibfnamefont {M.~A.}\ \bibnamefont {Caracanhas}}\ and\ \bibinfo {author} {\bibfnamefont {R.~G.}\ \bibnamefont {Pereira}},\ }\bibfield  {title} {\bibinfo {title} {Mobile impurities and orthogonality catastrophe in two-dimensional vortex lattices},\ }\href {https://doi.org/10.1103/PhysRevA.91.013608} {\bibfield  {journal} {\bibinfo  {journal} {Phys. Rev. A}\ }\textbf {\bibinfo {volume} {91}},\ \bibinfo {pages} {013608} (\bibinfo {year} {2015})}\BibitemShut {NoStop}%
\bibitem [{\citenamefont {Fukuhara}\ \emph {et~al.}(2013)\citenamefont {Fukuhara}, \citenamefont {Kantian}, \citenamefont {Endres}, \citenamefont {Cheneau}, \citenamefont {Schau{\ss}}, \citenamefont {Hild}, \citenamefont {Bellem}, \citenamefont {Schollw{\"o}ck}, \citenamefont {Giamarchi}, \citenamefont {Gross} \emph {et~al.}}]{Fukuhara_2013_quantum}%
  \BibitemOpen
  \bibfield  {author} {\bibinfo {author} {\bibfnamefont {T.}~\bibnamefont {Fukuhara}}, \bibinfo {author} {\bibfnamefont {A.}~\bibnamefont {Kantian}}, \bibinfo {author} {\bibfnamefont {M.}~\bibnamefont {Endres}}, \bibinfo {author} {\bibfnamefont {M.}~\bibnamefont {Cheneau}}, \bibinfo {author} {\bibfnamefont {P.}~\bibnamefont {Schau{\ss}}}, \bibinfo {author} {\bibfnamefont {S.}~\bibnamefont {Hild}}, \bibinfo {author} {\bibfnamefont {D.}~\bibnamefont {Bellem}}, \bibinfo {author} {\bibfnamefont {U.}~\bibnamefont {Schollw{\"o}ck}}, \bibinfo {author} {\bibfnamefont {T.}~\bibnamefont {Giamarchi}}, \bibinfo {author} {\bibfnamefont {C.}~\bibnamefont {Gross}}, \emph {et~al.},\ }\bibfield  {title} {\bibinfo {title} {Quantum dynamics of a mobile spin impurity},\ }\href@noop {} {\bibfield  {journal} {\bibinfo  {journal} {Nature Physics}\ }\textbf {\bibinfo {volume} {9}},\ \bibinfo {pages} {235} (\bibinfo {year} {2013})}\BibitemShut {NoStop}%
\bibitem [{\citenamefont {Catani}\ \emph {et~al.}(2012)\citenamefont {Catani}, \citenamefont {Lamporesi}, \citenamefont {Naik}, \citenamefont {Gring}, \citenamefont {Inguscio}, \citenamefont {Minardi}, \citenamefont {Kantian},\ and\ \citenamefont {Giamarchi}}]{Catani_2012_bose}%
  \BibitemOpen
  \bibfield  {author} {\bibinfo {author} {\bibfnamefont {J.}~\bibnamefont {Catani}}, \bibinfo {author} {\bibfnamefont {G.}~\bibnamefont {Lamporesi}}, \bibinfo {author} {\bibfnamefont {D.}~\bibnamefont {Naik}}, \bibinfo {author} {\bibfnamefont {M.}~\bibnamefont {Gring}}, \bibinfo {author} {\bibfnamefont {M.}~\bibnamefont {Inguscio}}, \bibinfo {author} {\bibfnamefont {F.}~\bibnamefont {Minardi}}, \bibinfo {author} {\bibfnamefont {A.}~\bibnamefont {Kantian}},\ and\ \bibinfo {author} {\bibfnamefont {T.}~\bibnamefont {Giamarchi}},\ }\bibfield  {title} {\bibinfo {title} {Quantum dynamics of impurities in a one-dimensional bose gas},\ }\href {https://doi.org/10.1103/PhysRevA.85.023623} {\bibfield  {journal} {\bibinfo  {journal} {Phys. Rev. A}\ }\textbf {\bibinfo {volume} {85}},\ \bibinfo {pages} {023623} (\bibinfo {year} {2012})}\BibitemShut {NoStop}%
\bibitem [{\citenamefont {Cavazos-Cavazos}\ \emph {et~al.}(2023)\citenamefont {Cavazos-Cavazos}, \citenamefont {Senaratne}, \citenamefont {Kafle},\ and\ \citenamefont {Hulet}}]{Cavazos_2023_thermal}%
  \BibitemOpen
  \bibfield  {author} {\bibinfo {author} {\bibfnamefont {D.}~\bibnamefont {Cavazos-Cavazos}}, \bibinfo {author} {\bibfnamefont {R.}~\bibnamefont {Senaratne}}, \bibinfo {author} {\bibfnamefont {A.}~\bibnamefont {Kafle}},\ and\ \bibinfo {author} {\bibfnamefont {R.~G.}\ \bibnamefont {Hulet}},\ }\bibfield  {title} {\bibinfo {title} {Thermal disruption of a luttinger liquid},\ }\href@noop {} {\bibfield  {journal} {\bibinfo  {journal} {Nature Communications}\ }\textbf {\bibinfo {volume} {14}},\ \bibinfo {pages} {3154} (\bibinfo {year} {2023})}\BibitemShut {NoStop}%
\bibitem [{\citenamefont {Kane}\ and\ \citenamefont {Fisher}(1992)}]{Kane_1992_transport}%
  \BibitemOpen
  \bibfield  {author} {\bibinfo {author} {\bibfnamefont {C.~L.}\ \bibnamefont {Kane}}\ and\ \bibinfo {author} {\bibfnamefont {M.~P.~A.}\ \bibnamefont {Fisher}},\ }\bibfield  {title} {\bibinfo {title} {Transport in a one-channel luttinger liquid},\ }\href {https://doi.org/10.1103/PhysRevLett.68.1220} {\bibfield  {journal} {\bibinfo  {journal} {Phys. Rev. Lett.}\ }\textbf {\bibinfo {volume} {68}},\ \bibinfo {pages} {1220} (\bibinfo {year} {1992})}\BibitemShut {NoStop}%
\bibitem [{\citenamefont {Mattsson}\ \emph {et~al.}(1997)\citenamefont {Mattsson}, \citenamefont {Eggert},\ and\ \citenamefont {Johannesson}}]{Mattsson_1997_boundaries}%
  \BibitemOpen
  \bibfield  {author} {\bibinfo {author} {\bibfnamefont {A.~E.}\ \bibnamefont {Mattsson}}, \bibinfo {author} {\bibfnamefont {S.}~\bibnamefont {Eggert}},\ and\ \bibinfo {author} {\bibfnamefont {H.}~\bibnamefont {Johannesson}},\ }\bibfield  {title} {\bibinfo {title} {Properties of a luttinger liquid with boundaries at finite temperature and size},\ }\href {https://doi.org/10.1103/PhysRevB.56.15615} {\bibfield  {journal} {\bibinfo  {journal} {Phys. Rev. B}\ }\textbf {\bibinfo {volume} {56}},\ \bibinfo {pages} {15615} (\bibinfo {year} {1997})}\BibitemShut {NoStop}%
\bibitem [{\citenamefont {J{\"u}ttner}\ \emph {et~al.}(1997)\citenamefont {J{\"u}ttner}, \citenamefont {Kl{\"u}mper},\ and\ \citenamefont {Suzuki}}]{Juttner_1997_exact}%
  \BibitemOpen
  \bibfield  {author} {\bibinfo {author} {\bibfnamefont {G.}~\bibnamefont {J{\"u}ttner}}, \bibinfo {author} {\bibfnamefont {A.}~\bibnamefont {Kl{\"u}mper}},\ and\ \bibinfo {author} {\bibfnamefont {J.}~\bibnamefont {Suzuki}},\ }\bibfield  {title} {\bibinfo {title} {Exact thermodynamics and luttinger liquid properties of the integrable tj model},\ }\href@noop {} {\bibfield  {journal} {\bibinfo  {journal} {Nuclear Physics B}\ }\textbf {\bibinfo {volume} {487}},\ \bibinfo {pages} {650} (\bibinfo {year} {1997})}\BibitemShut {NoStop}%
\bibitem [{\citenamefont {J{\"u}ttner}\ \emph {et~al.}(1998)\citenamefont {J{\"u}ttner}, \citenamefont {Kl{\"u}mper},\ and\ \citenamefont {Suzuki}}]{Juttner_1998_hubbard}%
  \BibitemOpen
  \bibfield  {author} {\bibinfo {author} {\bibfnamefont {G.}~\bibnamefont {J{\"u}ttner}}, \bibinfo {author} {\bibfnamefont {A.}~\bibnamefont {Kl{\"u}mper}},\ and\ \bibinfo {author} {\bibfnamefont {J.}~\bibnamefont {Suzuki}},\ }\bibfield  {title} {\bibinfo {title} {The hubbard chain at finite temperatures: ab initio calculations of tomonaga-luttinger liquid properties},\ }\href@noop {} {\bibfield  {journal} {\bibinfo  {journal} {Nuclear Physics B}\ }\textbf {\bibinfo {volume} {522}},\ \bibinfo {pages} {471} (\bibinfo {year} {1998})}\BibitemShut {NoStop}%
\bibitem [{\citenamefont {Le~Hur}(2006)}]{Le_Hur_2006_lifetime}%
  \BibitemOpen
  \bibfield  {author} {\bibinfo {author} {\bibfnamefont {K.}~\bibnamefont {Le~Hur}},\ }\bibfield  {title} {\bibinfo {title} {Electron lifetime in luttinger liquids},\ }\href {https://doi.org/10.1103/PhysRevB.74.165104} {\bibfield  {journal} {\bibinfo  {journal} {Phys. Rev. B}\ }\textbf {\bibinfo {volume} {74}},\ \bibinfo {pages} {165104} (\bibinfo {year} {2006})}\BibitemShut {NoStop}%
\bibitem [{\citenamefont {Horovitz}\ \emph {et~al.}(2013)\citenamefont {Horovitz}, \citenamefont {Giamarchi},\ and\ \citenamefont {Le~Doussal}}]{Horovitz_2013_Collective}%
  \BibitemOpen
  \bibfield  {author} {\bibinfo {author} {\bibfnamefont {B.}~\bibnamefont {Horovitz}}, \bibinfo {author} {\bibfnamefont {T.}~\bibnamefont {Giamarchi}},\ and\ \bibinfo {author} {\bibfnamefont {P.}~\bibnamefont {Le~Doussal}},\ }\bibfield  {title} {\bibinfo {title} {Phase transitions for a collective coordinate coupled to luttinger liquids},\ }\href {https://doi.org/10.1103/PhysRevLett.111.115302} {\bibfield  {journal} {\bibinfo  {journal} {Phys. Rev. Lett.}\ }\textbf {\bibinfo {volume} {111}},\ \bibinfo {pages} {115302} (\bibinfo {year} {2013})}\BibitemShut {NoStop}%
\bibitem [{\citenamefont {Kamenev}(2011)}]{Kamenev2011}%
  \BibitemOpen
  \bibfield  {author} {\bibinfo {author} {\bibfnamefont {A.}~\bibnamefont {Kamenev}},\ }\href@noop {} {\emph {\bibinfo {title} {Field theory of non-equilibrium systems}}}\ (\bibinfo  {publisher} {Cambridge University Press},\ \bibinfo {year} {2011})\BibitemShut {NoStop}%
\bibitem [{\citenamefont {Mahan}(1981)}]{mahan_book}%
  \BibitemOpen
  \bibfield  {author} {\bibinfo {author} {\bibfnamefont {G.~D.}\ \bibnamefont {Mahan}},\ }\href@noop {} {\emph {\bibinfo {title} {Many Particle Physics}}}\ (\bibinfo  {publisher} {Plenum},\ \bibinfo {address} {New York},\ \bibinfo {year} {1981})\BibitemShut {NoStop}%
\bibitem [{\citenamefont {Cazalilla}\ \emph {et~al.}(2011)\citenamefont {Cazalilla}, \citenamefont {Citro}, \citenamefont {Giamarchi}, \citenamefont {Orignac},\ and\ \citenamefont {Rigol}}]{Cazalilla_2011_bosons}%
  \BibitemOpen
  \bibfield  {author} {\bibinfo {author} {\bibfnamefont {M.~A.}\ \bibnamefont {Cazalilla}}, \bibinfo {author} {\bibfnamefont {R.}~\bibnamefont {Citro}}, \bibinfo {author} {\bibfnamefont {T.}~\bibnamefont {Giamarchi}}, \bibinfo {author} {\bibfnamefont {E.}~\bibnamefont {Orignac}},\ and\ \bibinfo {author} {\bibfnamefont {M.}~\bibnamefont {Rigol}},\ }\bibfield  {title} {\bibinfo {title} {One dimensional bosons: From condensed matter systems to ultracold gases},\ }\href {https://doi.org/10.1103/RevModPhys.83.1405} {\bibfield  {journal} {\bibinfo  {journal} {Rev. Mod. Phys.}\ }\textbf {\bibinfo {volume} {83}},\ \bibinfo {pages} {1405} (\bibinfo {year} {2011})}\BibitemShut {NoStop}%
\bibitem [{\citenamefont {Lieb}\ and\ \citenamefont {Liniger}(1963)}]{Lieb_1963_liniger}%
  \BibitemOpen
  \bibfield  {author} {\bibinfo {author} {\bibfnamefont {E.~H.}\ \bibnamefont {Lieb}}\ and\ \bibinfo {author} {\bibfnamefont {W.}~\bibnamefont {Liniger}},\ }\bibfield  {title} {\bibinfo {title} {Exact analysis of an interacting bose gas. i. the general solution and the ground state},\ }\href {https://doi.org/10.1103/PhysRev.130.1605} {\bibfield  {journal} {\bibinfo  {journal} {Phys. Rev.}\ }\textbf {\bibinfo {volume} {130}},\ \bibinfo {pages} {1605} (\bibinfo {year} {1963})}\BibitemShut {NoStop}%
\bibitem [{\citenamefont {Meinert}\ \emph {et~al.}(2017)\citenamefont {Meinert}, \citenamefont {Knap}, \citenamefont {Kirilov}, \citenamefont {Jag-Lauber}, \citenamefont {Zvonarev}, \citenamefont {Demler},\ and\ \citenamefont {N{\"a}gerl}}]{Meinert_2017_bloch}%
  \BibitemOpen
  \bibfield  {author} {\bibinfo {author} {\bibfnamefont {F.}~\bibnamefont {Meinert}}, \bibinfo {author} {\bibfnamefont {M.}~\bibnamefont {Knap}}, \bibinfo {author} {\bibfnamefont {E.}~\bibnamefont {Kirilov}}, \bibinfo {author} {\bibfnamefont {K.}~\bibnamefont {Jag-Lauber}}, \bibinfo {author} {\bibfnamefont {M.~B.}\ \bibnamefont {Zvonarev}}, \bibinfo {author} {\bibfnamefont {E.}~\bibnamefont {Demler}},\ and\ \bibinfo {author} {\bibfnamefont {H.-C.}\ \bibnamefont {N{\"a}gerl}},\ }\bibfield  {title} {\bibinfo {title} {Bloch oscillations in the absence of a lattice},\ }\href@noop {} {\bibfield  {journal} {\bibinfo  {journal} {Science}\ }\textbf {\bibinfo {volume} {356}},\ \bibinfo {pages} {945} (\bibinfo {year} {2017})}\BibitemShut {NoStop}%
\bibitem [{\citenamefont {Horvath}\ \emph {et~al.}(2026)\citenamefont {Horvath}, \citenamefont {Dhar}, \citenamefont {Wybo}, \citenamefont {Trypogeorgos}, \citenamefont {Guo}, \citenamefont {Zvonarev}, \citenamefont {Knap}, \citenamefont {Landini},\ and\ \citenamefont {Nägerl}}]{Horvath_2026_observing_impurity}%
  \BibitemOpen
  \bibfield  {author} {\bibinfo {author} {\bibfnamefont {M.}~\bibnamefont {Horvath}}, \bibinfo {author} {\bibfnamefont {S.}~\bibnamefont {Dhar}}, \bibinfo {author} {\bibfnamefont {E.}~\bibnamefont {Wybo}}, \bibinfo {author} {\bibfnamefont {D.}~\bibnamefont {Trypogeorgos}}, \bibinfo {author} {\bibfnamefont {Y.}~\bibnamefont {Guo}}, \bibinfo {author} {\bibfnamefont {M.}~\bibnamefont {Zvonarev}}, \bibinfo {author} {\bibfnamefont {M.}~\bibnamefont {Knap}}, \bibinfo {author} {\bibfnamefont {M.}~\bibnamefont {Landini}},\ and\ \bibinfo {author} {\bibfnamefont {H.-C.}\ \bibnamefont {Nägerl}},\ }\href {https://arxiv.org/abs/2602.12320} {\bibinfo {title} {Observing dissipationless flow of an impurity in a strongly repulsive quantum fluid}} (\bibinfo {year} {2026}),\ \Eprint {https://arxiv.org/abs/2602.12320} {arXiv:2602.12320 [cond-mat.quant-gas]} \BibitemShut {NoStop}%
\bibitem [{\citenamefont {Bakr}\ \emph {et~al.}(2009)\citenamefont {Bakr}, \citenamefont {Gillen}, \citenamefont {Peng}, \citenamefont {F{\"o}lling},\ and\ \citenamefont {Greiner}}]{Bakr_2009_microscope}%
  \BibitemOpen
  \bibfield  {author} {\bibinfo {author} {\bibfnamefont {W.~S.}\ \bibnamefont {Bakr}}, \bibinfo {author} {\bibfnamefont {J.~I.}\ \bibnamefont {Gillen}}, \bibinfo {author} {\bibfnamefont {A.}~\bibnamefont {Peng}}, \bibinfo {author} {\bibfnamefont {S.}~\bibnamefont {F{\"o}lling}},\ and\ \bibinfo {author} {\bibfnamefont {M.}~\bibnamefont {Greiner}},\ }\bibfield  {title} {\bibinfo {title} {A quantum gas microscope for detecting single atoms in a hubbard-regime optical lattice},\ }\href@noop {} {\bibfield  {journal} {\bibinfo  {journal} {Nature}\ }\textbf {\bibinfo {volume} {462}},\ \bibinfo {pages} {74} (\bibinfo {year} {2009})}\BibitemShut {NoStop}%
\bibitem [{\citenamefont {Bakr}\ \emph {et~al.}(2010)\citenamefont {Bakr}, \citenamefont {Peng}, \citenamefont {Tai}, \citenamefont {Ma}, \citenamefont {Simon}, \citenamefont {Gillen}, \citenamefont {Foelling}, \citenamefont {Pollet},\ and\ \citenamefont {Greiner}}]{Bakr_2010_probing}%
  \BibitemOpen
  \bibfield  {author} {\bibinfo {author} {\bibfnamefont {W.~S.}\ \bibnamefont {Bakr}}, \bibinfo {author} {\bibfnamefont {A.}~\bibnamefont {Peng}}, \bibinfo {author} {\bibfnamefont {M.~E.}\ \bibnamefont {Tai}}, \bibinfo {author} {\bibfnamefont {R.}~\bibnamefont {Ma}}, \bibinfo {author} {\bibfnamefont {J.}~\bibnamefont {Simon}}, \bibinfo {author} {\bibfnamefont {J.~I.}\ \bibnamefont {Gillen}}, \bibinfo {author} {\bibfnamefont {S.}~\bibnamefont {Foelling}}, \bibinfo {author} {\bibfnamefont {L.}~\bibnamefont {Pollet}},\ and\ \bibinfo {author} {\bibfnamefont {M.}~\bibnamefont {Greiner}},\ }\bibfield  {title} {\bibinfo {title} {Probing the superfluid--to--mott insulator transition at the single-atom level},\ }\href@noop {} {\bibfield  {journal} {\bibinfo  {journal} {Science}\ }\textbf {\bibinfo {volume} {329}},\ \bibinfo {pages} {547} (\bibinfo {year} {2010})}\BibitemShut {NoStop}%
\bibitem [{\citenamefont {Sherson}\ \emph {et~al.}(2010)\citenamefont {Sherson}, \citenamefont {Weitenberg}, \citenamefont {Endres}, \citenamefont {Cheneau}, \citenamefont {Bloch},\ and\ \citenamefont {Kuhr}}]{Sherson_2010_single}%
  \BibitemOpen
  \bibfield  {author} {\bibinfo {author} {\bibfnamefont {J.~F.}\ \bibnamefont {Sherson}}, \bibinfo {author} {\bibfnamefont {C.}~\bibnamefont {Weitenberg}}, \bibinfo {author} {\bibfnamefont {M.}~\bibnamefont {Endres}}, \bibinfo {author} {\bibfnamefont {M.}~\bibnamefont {Cheneau}}, \bibinfo {author} {\bibfnamefont {I.}~\bibnamefont {Bloch}},\ and\ \bibinfo {author} {\bibfnamefont {S.}~\bibnamefont {Kuhr}},\ }\bibfield  {title} {\bibinfo {title} {Single-atom-resolved fluorescence imaging of an atomic mott insulator},\ }\href@noop {} {\bibfield  {journal} {\bibinfo  {journal} {Nature}\ }\textbf {\bibinfo {volume} {467}},\ \bibinfo {pages} {68} (\bibinfo {year} {2010})}\BibitemShut {NoStop}%
\bibitem [{\citenamefont {Haller}\ \emph {et~al.}(2015)\citenamefont {Haller}, \citenamefont {Hudson}, \citenamefont {Kelly}, \citenamefont {Cotta}, \citenamefont {Peaudecerf}, \citenamefont {Bruce},\ and\ \citenamefont {Kuhr}}]{Haller_2015_single}%
  \BibitemOpen
  \bibfield  {author} {\bibinfo {author} {\bibfnamefont {E.}~\bibnamefont {Haller}}, \bibinfo {author} {\bibfnamefont {J.}~\bibnamefont {Hudson}}, \bibinfo {author} {\bibfnamefont {A.}~\bibnamefont {Kelly}}, \bibinfo {author} {\bibfnamefont {D.~A.}\ \bibnamefont {Cotta}}, \bibinfo {author} {\bibfnamefont {B.}~\bibnamefont {Peaudecerf}}, \bibinfo {author} {\bibfnamefont {G.~D.}\ \bibnamefont {Bruce}},\ and\ \bibinfo {author} {\bibfnamefont {S.}~\bibnamefont {Kuhr}},\ }\bibfield  {title} {\bibinfo {title} {Single-atom imaging of fermions in a quantum-gas microscope},\ }\href@noop {} {\bibfield  {journal} {\bibinfo  {journal} {Nature Physics}\ }\textbf {\bibinfo {volume} {11}},\ \bibinfo {pages} {738} (\bibinfo {year} {2015})}\BibitemShut {NoStop}%
\bibitem [{\citenamefont {Cheuk}\ \emph {et~al.}(2015)\citenamefont {Cheuk}, \citenamefont {Nichols}, \citenamefont {Okan}, \citenamefont {Gersdorf}, \citenamefont {Ramasesh}, \citenamefont {Bakr}, \citenamefont {Lompe},\ and\ \citenamefont {Zwierlein}}]{Cheuk_2015_fermions}%
  \BibitemOpen
  \bibfield  {author} {\bibinfo {author} {\bibfnamefont {L.~W.}\ \bibnamefont {Cheuk}}, \bibinfo {author} {\bibfnamefont {M.~A.}\ \bibnamefont {Nichols}}, \bibinfo {author} {\bibfnamefont {M.}~\bibnamefont {Okan}}, \bibinfo {author} {\bibfnamefont {T.}~\bibnamefont {Gersdorf}}, \bibinfo {author} {\bibfnamefont {V.~V.}\ \bibnamefont {Ramasesh}}, \bibinfo {author} {\bibfnamefont {W.~S.}\ \bibnamefont {Bakr}}, \bibinfo {author} {\bibfnamefont {T.}~\bibnamefont {Lompe}},\ and\ \bibinfo {author} {\bibfnamefont {M.~W.}\ \bibnamefont {Zwierlein}},\ }\bibfield  {title} {\bibinfo {title} {Quantum-gas microscope for fermionic atoms},\ }\href {https://doi.org/10.1103/PhysRevLett.114.193001} {\bibfield  {journal} {\bibinfo  {journal} {Phys. Rev. Lett.}\ }\textbf {\bibinfo {volume} {114}},\ \bibinfo {pages} {193001} (\bibinfo {year} {2015})}\BibitemShut {NoStop}%
\bibitem [{\citenamefont {Gross}\ and\ \citenamefont {Bakr}(2021)}]{Gross_2021_detection}%
  \BibitemOpen
  \bibfield  {author} {\bibinfo {author} {\bibfnamefont {C.}~\bibnamefont {Gross}}\ and\ \bibinfo {author} {\bibfnamefont {W.~S.}\ \bibnamefont {Bakr}},\ }\bibfield  {title} {\bibinfo {title} {Quantum gas microscopy for single atom and spin detection},\ }\href@noop {} {\bibfield  {journal} {\bibinfo  {journal} {Nature Physics}\ }\textbf {\bibinfo {volume} {17}},\ \bibinfo {pages} {1316} (\bibinfo {year} {2021})}\BibitemShut {NoStop}%
\bibitem [{\citenamefont {Knap}\ \emph {et~al.}(2013)\citenamefont {Knap}, \citenamefont {Kantian}, \citenamefont {Giamarchi}, \citenamefont {Bloch}, \citenamefont {Lukin},\ and\ \citenamefont {Demler}}]{Knap_2013_ramsey}%
  \BibitemOpen
  \bibfield  {author} {\bibinfo {author} {\bibfnamefont {M.}~\bibnamefont {Knap}}, \bibinfo {author} {\bibfnamefont {A.}~\bibnamefont {Kantian}}, \bibinfo {author} {\bibfnamefont {T.}~\bibnamefont {Giamarchi}}, \bibinfo {author} {\bibfnamefont {I.}~\bibnamefont {Bloch}}, \bibinfo {author} {\bibfnamefont {M.~D.}\ \bibnamefont {Lukin}},\ and\ \bibinfo {author} {\bibfnamefont {E.}~\bibnamefont {Demler}},\ }\bibfield  {title} {\bibinfo {title} {Probing real-space and time-resolved correlation functions with many-body ramsey interferometry},\ }\href {https://doi.org/10.1103/PhysRevLett.111.147205} {\bibfield  {journal} {\bibinfo  {journal} {Phys. Rev. Lett.}\ }\textbf {\bibinfo {volume} {111}},\ \bibinfo {pages} {147205} (\bibinfo {year} {2013})}\BibitemShut {NoStop}%
\bibitem [{\citenamefont {Jen}\ and\ \citenamefont {Wang}(2013)}]{Jen_2013_eit}%
  \BibitemOpen
  \bibfield  {author} {\bibinfo {author} {\bibfnamefont {H.~H.}\ \bibnamefont {Jen}}\ and\ \bibinfo {author} {\bibfnamefont {D.-W.}\ \bibnamefont {Wang}},\ }\bibfield  {title} {\bibinfo {title} {Theory of electromagnetically induced transparency in strongly correlated quantum gases},\ }\href {https://doi.org/10.1103/PhysRevA.87.061802} {\bibfield  {journal} {\bibinfo  {journal} {Phys. Rev. A}\ }\textbf {\bibinfo {volume} {87}},\ \bibinfo {pages} {061802(R)} (\bibinfo {year} {2013})}\BibitemShut {NoStop}%
\bibitem [{\citenamefont {Jen}\ and\ \citenamefont {Wang}(2014)}]{Jen_2014_extracting}%
  \BibitemOpen
  \bibfield  {author} {\bibinfo {author} {\bibfnamefont {H.}~\bibnamefont {Jen}}\ and\ \bibinfo {author} {\bibfnamefont {D.-W.}\ \bibnamefont {Wang}},\ }\bibfield  {title} {\bibinfo {title} {Extracting dynamical green’s function of ultracold quantum gases via electromagnetically induced transparency},\ }\href@noop {} {\bibfield  {journal} {\bibinfo  {journal} {Journal of the Optical Society of America B}\ }\textbf {\bibinfo {volume} {31}},\ \bibinfo {pages} {2931} (\bibinfo {year} {2014})}\BibitemShut {NoStop}%
\bibitem [{\citenamefont {Weitenberg}\ \emph {et~al.}(2026)\citenamefont {Weitenberg}, \citenamefont {Asteria}, \citenamefont {Carlsson}, \citenamefont {Bohrdt},\ and\ \citenamefont {Grusdt}}]{Weitenberg_2026_protocols}%
  \BibitemOpen
  \bibfield  {author} {\bibinfo {author} {\bibfnamefont {C.}~\bibnamefont {Weitenberg}}, \bibinfo {author} {\bibfnamefont {L.}~\bibnamefont {Asteria}}, \bibinfo {author} {\bibfnamefont {O.}~\bibnamefont {Carlsson}}, \bibinfo {author} {\bibfnamefont {A.}~\bibnamefont {Bohrdt}},\ and\ \bibinfo {author} {\bibfnamefont {F.}~\bibnamefont {Grusdt}},\ }\bibfield  {title} {\bibinfo {title} {Protocols for a many-body phase microscope: From coherences and $d$-wave superconductivity to green's functions},\ }\href {https://doi.org/10.1103/99mf-c8tv} {\bibfield  {journal} {\bibinfo  {journal} {PRX Quantum}\ }\textbf {\bibinfo {volume} {7}},\ \bibinfo {pages} {033068} (\bibinfo {year} {2026})}\BibitemShut {NoStop}%
\end{thebibliography}%

\newpage
\clearpage
\appendix

\onecolumngrid

\section{Derivation of the real-space Green's function} \label{SM:sec:GF_finite_T_LCE}

We construct an estimate of the single-particle Green's function of the impurity in real space:
\begin{equation}
\bra{0} \hat d(y) e^{-i\hat H t} \hat d^{\dag}(x) \ket{0}=\frac{1}{L}\sum_{k_1,k_2} e^{i(k_2 y -k_1 x)}\bra{0} \hat d_{k_2}(t) \hat d^{\dag}_{k_1}(0) \rangle\ket{0}=\frac{1}{L}\sum_{k_1,k_2} e^{i(k_2 y -k_1 x)}\langle \varphi_{k_2,+}(t) \varphi^{*}_{k_1,+}(0) \rangle,
\end{equation}
where the average is taken over the action obtained by summing:
\begin{align}\label{S:Eq:S_bath}
&S_{bath}=\sum_{q\neq 0} \int_{0}^{+\infty} dt \int_{0}^{+\infty} dt' \sum_{\alpha,\beta = cl, qu} \phi^*_{q,\alpha}(t)\mathcal{G}^{-1}_{\alpha,\beta}\phi_{q,\beta}(t'),\\
&\mathcal{G}^{-1}=\begin{pmatrix}
0 & G_{A,bath}^{-1}(t,t') \\
 G_{R,bath}^{-1}(t,t')& i\frac{1+w_q}{1-w_q}\delta(t)\delta(t')
\end{pmatrix},
\end{align}
\begin{align}\label{S:Eq:S_imp}
&S_{imp}=\sum_{k}  \int_{0}^{+\infty} dt \int_{0}^{+\infty} dt' \sum_{\alpha,\beta = cl, qu} \varphi^*_{k,\alpha}(t)\mathcal{G}^{-1}_{\alpha,\beta}\varphi_{k,\beta}(t'),\\
&\mathcal{G}^{-1}=\begin{pmatrix}
0 &  G_{A,imp}^{-1}(t,t')\\
G_{R,imp}^{-1}(t,t')& i\frac{1+u_k}{1-u_k}\delta(t)\delta(t')
\end{pmatrix},
\end{align}
\begin{align}\label{S:Eq:S_int}
&S_{int}=\int dt\,\, \left[-H_{int}\left(\phi_{+}(t),\varphi_+(t)\right)+H_{int}\left(\phi_{-}(t),\varphi_-(t)\right) \right]=\\
&=-\frac{g\sqrt{K}}{L}\sum_k \sum_{p\neq 0}V(p)\int_{-\infty}^{+\infty} dt\, \Biggl\{\varphi^*_{k,+}(t) \varphi_{k+p,+}(t)\left[\phi^*_{p,+}(t)+\phi_{-p,+}(t) \right]-\varphi^*_{k,-}(t) \varphi_{k+p,-}(t)\left[\phi^*_{p,-}(t)+\phi_{-p,-}(t) \right] \Biggr\}, 
\end{align}
where $V(p)=\sqrt{\frac{L|p|}{2\pi}}e^{-\alpha\frac{|p|}{2}}$, together with the vacuum conditions $u_k=0$ $\forall k$ and $w_q=e^{-\beta\omega_q}$ $\forall q$. The free correlators read:
\begin{align}
& \langle \varphi_{k,\pm}(t)\varphi^*_{k',\mp}(t')\rangle_{S_0}= \delta_{k,k'}e^{-i\epsilon_k(t-t')}\frac{F(\epsilon_k)\mp 1}{2},\label{SM:Eq:corr1}\\
&\langle  \varphi_{k,\pm}(t)\varphi^*_{k',\pm}(t') \rangle_{S_0} =\theta(t-t')\delta_{k,k'}e^{-i\epsilon_k(t-t')}\frac{F(\epsilon_k)\pm 1}{2}+\theta(t'-t)\delta_{k,k'}e^{-i\epsilon_k(t-t')}\frac{F(\epsilon_k)\mp 1}{2}\\
& \langle \phi_{q,\pm}(t)\phi^*_{q',\mp}(t')\rangle_{S_0}= \delta_{q,q'}e^{-i\omega_q(t-t')}\frac{F(\omega_q)\mp 1}{2},\label{SM:Eq:corr3}\\
&\langle  \phi_{q,\pm}(t)\phi^*_{q',\pm}(t') \rangle_{S_0} =\theta(t-t')\delta_{q,q'}e^{-i\omega_q(t-t')}\frac{F(\omega_q)\pm 1}{2}+\theta(t'-t)\delta_{q,q'}e^{-i\omega_q(t-t')}\frac{F(\omega_q)\mp 1}{2},\label{SM:Eq:corr4}
\end{align}
where $F(\epsilon)=1+2n_B (\epsilon)$ and $n_B(\epsilon)=(e^{\beta\epsilon}-1)^{-1}$ is the Bose-Einstein distribution function. The finite-temperature condition is applied solely to the bath degrees of freedom.

An expansion to second order yields the result:
\begin{equation}
\begin{split}
&\bra{0} \hat d(y) e^{-i\hat H t} \hat d^{\dag}(x) \ket{0}= \frac{1}{L}\sum_{k_1,k_2}e^{i(k_2 y-k_1 x)} \delta_{k_1,k_2} e^{-i\epsilon_{k_1}t} \left[1-\int du \Biggl(R^{(a)}(u;k_1,T)+R^{(e)}(u;k_1,T)\Biggr)\frac{1-iut-e^{-iut}}{u^2} \right]\approx\\
&\approx \int \frac{dk}{2\pi}e^{ik(y-x)-i\epsilon_k t-\int du [R^{(a)}(u;k,T)+R^{(e)}(u;k,T)]\frac{1-iut-e^{-iut}}{u^2}},
\label{S:Eq:GF_T}
\end{split}
\end{equation} 
where $R^{(a/e)}(u;k,T)$ is defined as:
\begin{align}
&R^{(a)}(u;k,T)=\frac{g^2 K}{(2\pi)^2}\int dp\,\, |p| e^{-\alpha|p|}n_B(\omega_p;T)\,\,\delta(u-\epsilon_{k+p}+\omega_p+\epsilon_k),\label{S:Eq:Ra}\\
&R^{(e)}(u;k,T)=\frac{g^2 K}{(2\pi)^2}\int dp\,\, |p| e^{-\alpha|p|}\Bigl(n_B(\omega_p;T)+1\Bigr)\,\,\delta(u-\epsilon_{k+p}-\omega_p+\epsilon_k),\label{S:Eq:Re}
\end{align}
and in the second equality we have reexponentiated the term in square brackets.

\section{Review of the zero-temperature results for the momentum-space single-particle Green's function}\label{SM:sec:zero_T}

Here, we present the large-time expressions of the single-particle Green's function in the distinct dynamical regimes, in order to provide the zero-temperature expressions that finite-temperature corrections systematically alter. We start by realizing that the $T=0$ momentum-space single-particle Green's function can be read off from Eq.~\eqref{S:Eq:GF_T} as:
\begin{equation}
    G(k,t;\beta=+\infty)= e^{-i\epsilon_k t -\int du R^{(e)}(u;k, T=0)\frac{1-iut-e^{-iut}}{u^2}},
\end{equation}
whose modulus takes the form:
\begin{equation}\label{S:Eq:zero_T_GF}
    |G(k,t;\beta=+\infty)| = \exp\left\{-\frac{g^2 K}{(2\pi)^2} \int dp\, |p|e^{-\alpha |p|} \frac{1-\cos\left[(\epsilon_{k+p} + \omega_p -\epsilon_k)t \right]}{(\epsilon_{k+p} + \omega_p -\epsilon_k)^2} \right\},
\end{equation}
where we assume a quadratic impurity dispersion relation, i.e., $\epsilon_k = t_i k^2$, together with the linear dispersion relation $\omega_q = v_s |q|$ of the sound modes of the Luttinger liquid theory. The large-time limit of Eq.~\eqref{S:Eq:zero_T_GF} is presented here below in the distinct dynamical regimes $0\leq k < v_s/(2 t_i)$ and $k>v_s/(2 t_i)$.

In the first case, i.e., $0\leq k < v_s/(2 t_i)$, one obtains the following power-law decay:
\begin{equation}
    |G(k,t;\beta=+\infty)|\approx \left( \frac{2t_i k +v_s}{\alpha}t \right)^{-\frac{g^2 K}{(2\pi)^2 (2t_i k + v_s)^2}} \left( \frac{v_s - 2t_i k }{\alpha}t \right)^{-\frac{g^2 K}{(2\pi)^2 ( v_s -2t_i k)^2}},
\end{equation}
associated with an instance of the Anderson's orthogonality catastrophe in the one-dimensional bath.
In the second case, namely $k>v_s/(2t_i)$, the large-time estimate gives instead a leading linear-in-time term in the argument of the exponential of the expression of the zero-temperature Green's function, thus resulting in the exponential decay:
\begin{equation}
    |G(k,t;\beta=+\infty)|\approx \exp\left\{-\,\frac{g^{2}K}{(2\pi)^{2}}\;
\frac{\pi\,e^{-\alpha\frac{2 t_i k-v_s}{ t_i}}}{ t_i}\;t\right\},
\end{equation}
that is associated with the onset of quasi-particle behavior.

\section{Asymptotic large-time behavior of the momentum-space Green's function at finite temperature}\label{S:G_k_large_t}

From Eq.\eqref{S:Eq:GF_T}, one can read off the correction of the modulus of the momentum-space Green's function due to a finite temperature, which takes the form:
\begin{equation} \label{Eq:concrete_formula_general}
  \frac{|G(k,t;\beta)|}{|G(k,t;+\infty)|} = e^{-\frac{g^2 K}{(2\pi)^2}\int dp \, |p| e^{-\alpha |p|} n_B (\omega_p;T)\left\{\frac{1-\cos[(\epsilon_{k+p}-\epsilon_k -\omega_p)t]}{(\epsilon_{k+p}-\epsilon_k -\omega_p)^2} + \frac{1-\cos[(\epsilon_{k+p}-\epsilon_k +\omega_p)t]}{(\epsilon_{k+p}-\epsilon_k +\omega_p)^2} \right\}}.
\end{equation}
We analyze the large-time behavior of the above expression in the case of a quadratic dispersion relation, defined by the relation $\epsilon_k = t_i k^2$.

\subsection{The case of a quadratic dispersion relation}

If the dispersion relation is quadratic, Eq.~\eqref{Eq:concrete_formula_general} reads:
\begin{equation} \label{Eq:total_result}
\begin{split}
    & \frac{|G(k,t;\beta)|}{|G(k,t;+\infty)|} = \\
    & =\exp \Biggl\{{-\frac{g^2 K}{(2\pi)^2}\int dp \, |p| e^{-\alpha |p|} n_B (\omega_p;T)\left\{\frac{1-\cos[(t_i(k+p)^2-t_i k^2 -\omega_p)t]}{(t_i(k+p)^2-t_i k^2 -\omega_p)^2} + \frac{1-\cos[(t_i(k+p)^2-t_i k^2 +\omega_p)t]}{(t_i(k+p)^2-t_i k^2 +\omega_p)^2} \right\}}\Biggr\}=\\
    &=\exp \Biggl\{-\frac{g^2 K}{(2\pi)^2}(I_1 +I_2)\Biggr\}.
     \end{split}
\end{equation}

The above expression is extremely challenging, yet, it is possible to make analytical progress by noticing that the integrals above will be dominated by the behavior of the function close to the poles of the arguments of the cosine function. Let us discuss separately the two integrals. The first one can be manipulated as:
\begin{equation}\label{S:Eq:I1}
\begin{split}
   & I_1= \int dp \, |p| e^{-\alpha |p|} n_B (\omega_p;T)\frac{1-\cos[(t_i(k+p)^2-t_i k^2 -\omega_p)t]}{[t_i(k+p)^2-t_i k^2 -\omega_p]^2}= \\
 &=\int _0^{+\infty}dp\, p\, e^{-\alpha p} n_B (\omega_p;T)\frac{1-\cos[(t_i(k+p)^2-t_i k^2 -v_s p)t]}{[t_i(k+p)^2-t_i k^2 -v_s p]^2} + \int_0^{+\infty} dp \, p\, e^{-\alpha p} n_B (\omega_p;T)\frac{1-\cos[(t_i(k-p)^2-t_i k^2 -v_s p)t]}{[t_i(k-p)^2-t_i k^2 -v_s p]^2}=\\
 &= \int _0^{+\infty}dp\, p\, e^{-\alpha p} n_B (\omega_p;T)\frac{1-\cos[(2t_i k -v_s +t_i p)pt]}{[(2t_i k -v_s +t_i p)p]^2} + \int_0^{+\infty} dp \, p\, e^{-\alpha p} n_B (\omega_p;T)\frac{1-\cos[(t_i p- 2 t_i k -v_s )pt]}{[(t_i p- 2 t_i k -v_s )p]^2}. 
    \end{split}
\end{equation}

Analogously, the second integral can be rewritten as:
\begin{equation}\label{S:Eq:I2}
\begin{split}
   & I_2= \int dp \, |p| e^{-\alpha |p|} n_B (\omega_p;T)\frac{1-\cos[(t_i(k+p)^2-t_i k^2 +\omega_p)t]}{[t_i(k+p)^2-t_i k^2 +\omega_p]^2}= \\
 &=\int _0^{+\infty}dp\, p\, e^{-\alpha p} n_B (\omega_p;T)\frac{1-\cos[(t_i(k+p)^2-t_i k^2 +v_s p)t]}{[t_i(k+p)^2-t_i k^2 + v_s p]^2} + \int_0^{+\infty} dp \, p\, e^{-\alpha p} n_B (\omega_p;T)\frac{1-\cos[(t_i(k-p)^2-t_i k^2 +v_s p)t]}{[t_i(k-p)^2-t_i k^2 +v_s p]^2}=\\
 &= \int _0^{+\infty}dp\, p\, e^{-\alpha p} n_B (\omega_p;T)\frac{1-\cos[(2t_i k +v_s +t_i p)pt]}{[(2t_i k + v_s +t_i p)p]^2} + \int_0^{+\infty} dp \, p\, e^{-\alpha p} n_B (\omega_p;T)\frac{1-\cos[(t_i p- 2 t_i k +v_s )pt]}{[(t_i p- 2 t_i k + v_s )p]^2}. 
    \end{split}
\end{equation}
To this end, it is now necessary to distinguish between two cases, namely, $0 \leq k < v_s/(2 t_i)$ or $k>v_s/(2t_i)$, as well as present a quick discussion of the special point $k=v_s/(2t_i)$.

\subsubsection{The case $0 \leq k < v_s/(2 t_i)$}

In the present case, it is possible to notice that the first integrand in the final expression of $I_1$ is dominated by the poles of the argument of the cosine, which are given by $p_1^* = 0$ and $p_2^* =(v_s - 2t_i k)/t_i$, whereas the same consideration for the second integrand leads to the identification of the poles $p_3^* = p_1^* = 0$ and $p_4^* = (2t_i k+v_s)/t_i$. Notice how these poles fall either inside the domain of integration or at its left boundary. In a similar fashion, in the case of $I_2$, it is possible to notice that the only relevant poles of the cosine arguments of the integrands are the ones located at $p^*_5=p^*_6=0$.

To proceed with the large-time estimate, we approximate the full integrals in $I_1$ and $I_2$ by restricting the integration domain to the neighborhoods of the poles, by linearizing the argument of the cosine around the poles and then, after a change of variable, by using the delta-function representation:
\begin{equation}
    \lim_{t\rightarrow +\infty} \frac{1-\cos(tx)}{\pi t x^2} = \delta(x).
\end{equation}

The estimate of $I_1$ proceeds as follows:
\begin{equation}
    \begin{split}
        & I_1 \approx \int _{p \approx 0^+}dp\, p\, e^{-\alpha p} n_B (\omega_p;T)\frac{1-\cos[(2t_i k -v_s )pt]}{[(2t_i k -v_s)p]^2} + \int_{p \approx \frac{v_s -2t_i k}{t_i}} dp \, p\, e^{-\alpha p} n_B (\omega_p;T)\frac{1-\cos[(v_s - 2t_i k)( p-\frac{v_s -2t_i k}{t_i})t]}{\left[(v_s - 2t_i k)( p-\frac{v_s -2t_i k}{t_i})\right]^2}+\\
        & +\int _{p \approx 0^+}dp\, p\, e^{-\alpha p} n_B (\omega_p;T)\frac{1-\cos[(2t_i k +v_s )pt]}{[(2t_i k +v_s)p]^2} + \int_{p \approx \frac{v_s +2t_i k}{t_i}} dp \, p\, e^{-\alpha p} n_B (\omega_p;T)\frac{1-\cos[(v_s + 2t_i k)( p-\frac{v_s + 2t_i k}{t_i})t]}{\left[(v_s + 2t_i k)( p-\frac{v_s +2t_i k}{t_i})\right]^2}\approx\\
        &\approx \frac{\pi t}{2\beta v_s}\left( \frac{1}{v_s -2 t_i k} + \frac{1}{v_s + 2 t_i k}\right) + \frac{\pi t}{t_i}\frac{e^{-\alpha\frac{v_s -2t_i k}{t_i}}}{e^{\beta v_s\frac{v_s -2t_i k}{t_i}} -1}+\frac{\pi t}{t_i}\frac{e^{-\alpha\frac{v_s +2t_i k}{t_i}}}{e^{\beta v_s\frac{v_s + 2t_i k}{t_i}} -1}.
    \end{split}
\end{equation}
Analogously, one may estimate $I_2$ as:
\begin{equation}
    \begin{split}
        & I_2 \approx \int _{p \approx 0^+}dp\, p\, e^{-\alpha p} n_B (\omega_p;T)\frac{1-\cos[(2t_i k -v_s )pt]}{[(2t_i k -v_s)p]^2}  +\int _{p \approx 0^+}dp\, p\, e^{-\alpha p} n_B (\omega_p;T)\frac{1-\cos[(2t_i k +v_s )pt]}{[(2t_i k +v_s)p]^2} \approx\\
        &\approx \frac{\pi t}{2\beta v_s}\left( \frac{1}{v_s -2 t_i k} + \frac{1}{v_s + 2 t_i k}\right).
    \end{split}
\end{equation}

Plugging the obtained asymptotic estimates for $I_1$ and $I_2$ into Eq.~\eqref{Eq:total_result}, one can write:
\begin{equation}\label{S:Eq:k<}
\begin{split}
    \frac{|G(k,t;\beta)|}{|G(k,t;+\infty)|} \approx\exp \Biggl\{-\frac{g^2 K}{(2\pi)^2} \left[\frac{\pi t}{\beta v_s}\left( \frac{1}{v_s -2 t_i k} + \frac{1}{v_s + 2 t_i k}\right) + \frac{\pi t}{t_i}\frac{e^{-\alpha\frac{v_s -2t_i k}{t_i}}}{e^{\beta v_s\frac{v_s -2t_i k}{t_i}} -1}+\frac{\pi t}{t_i}\frac{e^{-\alpha\frac{v_s +2t_i k}{t_i}}}{e^{\beta v_s\frac{v_s + 2t_i k}{t_i}} -1} \right]\Biggr\}.
    \end{split}
\end{equation}

\subsubsection{The case  $k > v_s/(2 t_i)$}

The case $k> v_s/(2t_i)$ can be treated in the same way as the opposed scenario by identifying the poles of the argument of the cosine functions appearing in the expressions of $I_1$ and $I_2$ such that they fall within the domain of integration. In this way, it is immediate to realize that $I_1$ is dominated by the behavior of the first integrand at the pole $p_1^* =0$ and of the second integrand at the poles $p_2^* =0$ and $p_3^* = (2t_i k + v_s)/t_i$, while $I_2$ is dominated by the behavior of the first integrand at the pole $p_4^*=0$ and of the second integrand at the poles $p_5^*=0$ and $p_6^* =\frac{2t_i k - v_s}{t_i}$.

The estimate of $I_1$ can be carried out as follows:
\begin{equation}
    \begin{split}
        & I_1 \approx \int _{p \approx 0^+}dp\, p\, e^{-\alpha p} n_B (\omega_p;T)\frac{1-\cos[(2t_i k -v_s )pt]}{[(2t_i k -v_s)p]^2} +\int _{p \approx 0^+}dp\, p\, e^{-\alpha p} n_B (\omega_p;T)\frac{1-\cos[(2t_i k +v_s )pt]}{[(2t_i k +v_s)p]^2} +\\
        &  + \int_{p \approx \frac{v_s +2t_i k}{t_i}} dp \, p\, e^{-\alpha p} n_B (\omega_p;T)\frac{1-\cos[(v_s + 2t_i k)( p-\frac{v_s + 2t_i k}{t_i})t]}{\left[(v_s + 2t_i k)( p-\frac{v_s +2t_i k}{t_i})\right]^2}\approx\\
        &\approx \frac{\pi t}{2\beta v_s}\left( \frac{1}{2 t_i k -v_s} + \frac{1}{v_s + 2 t_i k}\right)  +\frac{\pi t}{t_i}\frac{e^{-\alpha\frac{v_s +2t_i k}{t_i}}}{e^{\beta v_s\frac{v_s + 2t_i k}{t_i}} -1}.
    \end{split}
\end{equation}
Similarly, the estimate of $I_2$ is performed as:
\begin{equation}
    \begin{split}
        & I_2 \approx \int _{p \approx 0^+}dp\, p\, e^{-\alpha p} n_B (\omega_p;T)\frac{1-\cos[(2t_i k -v_s )pt]}{[(2t_i k -v_s)p]^2}  +\int _{p \approx 0^+}dp\, p\, e^{-\alpha p} n_B (\omega_p;T)\frac{1-\cos[(2t_i k +v_s )pt]}{[(2t_i k +v_s)p]^2}+\\
        & + \int _{p \approx \frac{2t_i k-v_s}{t_i}}dp\, p\, e^{-\alpha p} n_B (\omega_p;T)\frac{1-\cos[(2t_i k -v_s )\left(p - \frac{2t_i k-v_s}{t_i} \right)t]}{\left[(2t_i k -v_s )\left(p - \frac{2t_i k-v_s}{t_i} \right)\right]^2}  \approx\\
        &\approx \frac{\pi t}{2\beta v_s}\left( \frac{1}{2 t_i k -v_s} + \frac{1}{v_s + 2 t_i k}\right)+\frac{\pi t}{t_i}\frac{e^{-\alpha\frac{2t_i k - v_s}{t_i}}}{e^{\beta v_s\frac{2t_i k - v_s}{t_i}} -1}.
    \end{split}
\end{equation}
The end result for the ratio $|G(k,t;\beta)|/|G(k,t;+\infty)|$ reads therefore:
\begin{equation}\label{S:Eq:k>}
\begin{split}
    \frac{|G(k,t;\beta)|}{|G(k,t;+\infty)|} \approx\exp \Biggl\{-\frac{g^2 K}{(2\pi)^2} \left[\frac{\pi t}{\beta v_s}\left( \frac{1}{2 t_i k -v_s} + \frac{1}{v_s + 2 t_i k}\right) + \frac{\pi t}{t_i}\frac{e^{-\alpha\frac{2t_i k - v_s}{t_i}}}{e^{\beta v_s\frac{2t_i k - v_s}{t_i}} -1}+\frac{\pi t}{t_i}\frac{e^{-\alpha\frac{v_s +2t_i k}{t_i}}}{e^{\beta v_s\frac{v_s + 2t_i k}{t_i}} -1} \right]\Biggr\}.
    \end{split}
\end{equation}

\subsubsection{The case  $k = v_s/(2 t_i)$}

Here, we discuss the exceptional behavior occurring for $k=v_s/(2t_i)$. To this end, we observe that:
\begin{equation}\label{S:Eq:delta_rep}
  \lim_{t\rightarrow +\infty}  \frac{3}{2}\sqrt{\frac{t_i}{2 \pi t^3}}\frac{1 - \cos(t_i p^2 t)}{(t_i p^2)^2}= \delta(p).
\end{equation}
By means of the identity presented in Eq.~\eqref{S:Eq:delta_rep}, we manipulate the first term in the last expression of Eq.~\eqref{S:Eq:I1} in the large-time limit as follows:
\begin{equation}\label{S:Eq:result_special}
\begin{split}
    \int _0^{+\infty}dp\, p\, e^{-\alpha p} n_B (\omega_p;T)\frac{1-\cos[(2t_i k -v_s +t_i p)pt]}{[(2t_i k -v_s +t_i p)p]^2} \Biggr|_{k=k_c} = \int _0^{+\infty}dp\, p\, e^{-\alpha p} n_B (\omega_p;T)\frac{1-\cos[t_i p^2 t]}{(t_i p^2)^2}\approx \frac{2}{3}\sqrt{\frac{2\pi t^3}{t_i}} \frac{1}{2\beta v_s}.
    \end{split}
\end{equation}
where we considered the large-time limit in the last step of the calculation. An analogous manipulation of the second term in the last expression of Eq.~\eqref{S:Eq:I2} leads to the same result as the one presented in Eq.~\eqref{S:Eq:result_special}. Thus, after treating the remaining terms as done in the other parameter regimes, the leading finite-temperature correction to the single-particle Green's function reads in this case:  \begin{equation}\label{S:Eq:finite_T_corr_kc}
\begin{split}
    \frac{|G(k,t;\beta)|}{|G(k,t;+\infty)|} \approx\exp \Biggl\{-\frac{g^2 K}{(2\pi)^2} \left[\frac{2}{3\beta v_s} \sqrt{\frac{2\pi}{t_i}} t^{3/2} + \frac{\pi t}{2\beta v_s^2}+ \frac{\pi t}{t_i} \frac{e^{-2\alpha v_s/t_i}}{e^{2\beta v_s^2 /t_i} - 1} \right]\Biggr\}.
    \end{split}
\end{equation}
Eq.~\eqref{S:Eq:finite_T_corr_kc} thus displays an unexpected, novel leading-order term in the argument of the exponent, that scales non-analytically as $t^{3/2}$, rather than linearly with time $t$. This feature, associated with the exceptional critical momentum value $k_c$, is responsible for the divergence of the temperature-induced coherence time $\tau$ while $k\rightarrow k_c^{\pm}$.

\section{Convergence of the asymptotic formulas}\label{S:convergence}

\begin{figure}[h]
\includegraphics[width=\textwidth]{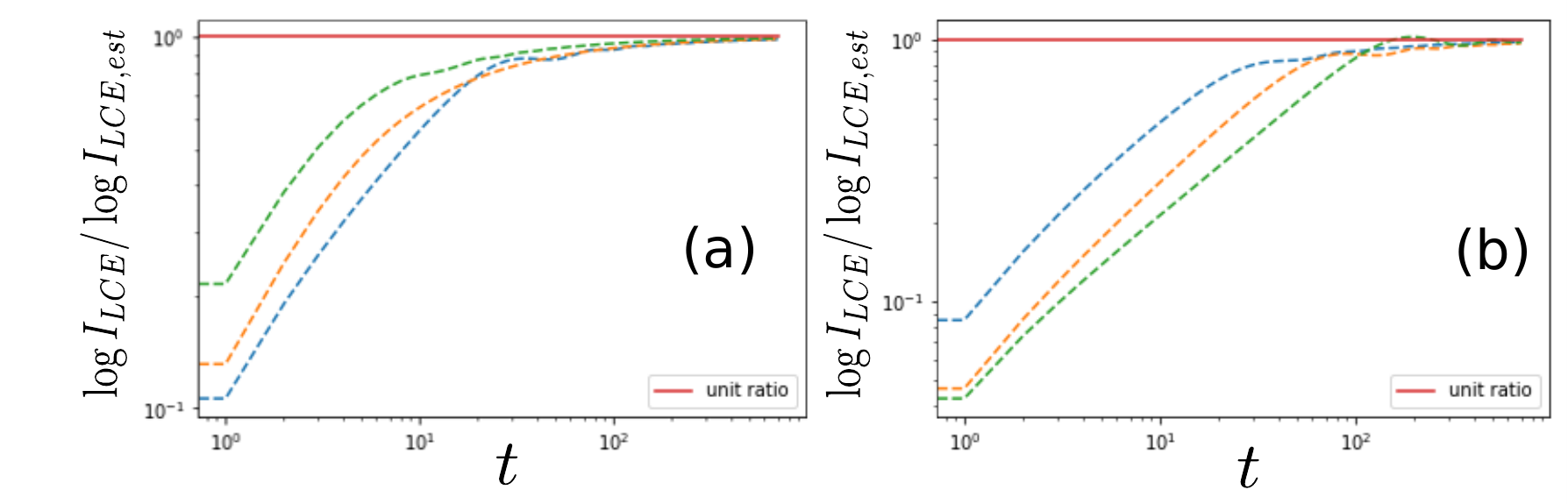}
\caption{Ratio $\log I_{LCE}/\log I_{LCE,est} $ as a function of time $t$ for parameter values $k = 1.0,
g = 0.5,
K = 1.2,
t_i = 0.3,
v_s = 1.0,
\alpha = 0.5, \beta=2.0$ (blue curve), $k = 0.5,
g = 0.75,
K = 1.1,
t_i = 0.25,
v_s = 1.1,
\alpha = 0.5, \beta = 3.0
$ (orange curve) and $k = 0.25,
g = 0.25,
K = 1.3,
t_i = 0.5,
v_s = 1.25,
\alpha = 0.5, \beta = 1.5$ (green curve), respectively, compared to the value $1$ (continuous red curve).}
\label{Fig:check_tot}
\end{figure}

In the present section, we present additional details on the convergence of the l.h.s. of Eqs.~\eqref{S:Eq:k<}-\eqref{S:Eq:k>}, derived with a linked-cluster expansion and displayed explicitly in Eq.~\eqref{Eq:concrete_formula_general}, to the analytical large-time approximations that are shown on the r.h.s. of the very same equations. Indeed, by denoting the full expression of Eq.~\eqref{Eq:concrete_formula_general} as $I_{LCE}$ and the r.h.s. of Eqs.~\eqref{S:Eq:k<}-\eqref{S:Eq:k>} as $I_{LCE,est}$, namely its asymptotic long-time estimate, we observe numerically in Fig.~\ref{Fig:check_tot} that the ratio $\log I_{LCE}/\log I_{LCE,est}$ displays a behavior that is consistent with its convergence to $1$ for all explored parameter values. This highlights the precise meaning of the asymptotic approximations displayed in the main text of the manuscript, which consist indeed in the evaluation of the large-time leading term in the argument of the exponent of Eq.~\eqref{Eq:concrete_formula_general}.

\end{document}